\pdfoutput=1
\documentclass[journal,onecolumn,11pt]{IEEEtran}

\usepackage[utf8]{inputenc}
\usepackage[T1]{fontenc}
\usepackage{amsmath,amssymb,amsthm}
\usepackage{mathtools}
\usepackage{booktabs}
\usepackage{array}
\usepackage{multirow}
\usepackage{graphicx}
\usepackage{xcolor}
\usepackage{tikz}
\usetikzlibrary{arrows.meta}
\usepackage{cite}
\usepackage[hidelinks]{hyperref}
\usepackage{orcidlink}

\newtheorem{theorem}{Theorem}
\newtheorem{lemma}[theorem]{Lemma}
\newtheorem{proposition}[theorem]{Proposition}
\newtheorem{corollary}[theorem]{Corollary}
\newtheorem{conjecture}[theorem]{Conjecture}
\theoremstyle{definition}

\newcommand{\F}{\mathbb{F}}

\newcommand{\Fix}{\operatorname{Fix}}
\newcommand{\Img}{\operatorname{Im}}
\newcommand{\rank}{\operatorname{rank}}
\newcommand{\Tr}{\operatorname{Tr}}
\newcommand{\GL}{\operatorname{GL}}
\newcommand{\GammaL}{\operatorname{\Gamma L}}
\newcommand{\Herm}{\operatorname{Herm}}
\newcommand{\PG}{\operatorname{PG}}
\newcommand{\spn}{\operatorname{span}}
\newcommand{\artifact}{\url{https://github.com/peacker/apn10-self-equivalences}}

\begin{document}

\title{Quadratic APN Functions of $\mathbb{F}_2^{10}$ with a\\ Self-Equivalence of Prime Order}

\author{Emanuele~Bellini\,\orcidlink{0000-0002-2349-0247}%
\thanks{E.~Bellini is with the Technology Innovation Institute, Abu Dhabi, United Arab Emirates (e-mail: emanuele.bellini@tii.ae).}%
\thanks{This work has been submitted to the IEEE for possible publication. Copyright may be transferred without notice, after which this version may no longer be accessible.}}

\markboth{Preprint, submitted to IEEE Transactions on Information Theory}%
{Bellini: Quadratic APN Functions of $\mathbb{F}_2^{10}$ with a Self-Equivalence of Prime Order}

\maketitle

\begin{abstract}
Almost perfect nonlinear (APN) functions of $\mathbb{F}_2^n$ have the lowest possible differential uniformity.
The only known APN permutation in even dimension, in dimension six, is equivalent to a quadratic APN function with a nontrivial linear self-equivalence; no APN permutation is known for even $n\ge 8$.
In a published collection of about 3.8 million quadratic APN classes of dimension eight, at most 0.34\% have a nontrivial linear self-equivalence.
In dimension ten, the 25 literature classes considered here all have one.
We study all quadratic APN functions of $\mathbb{F}_2^{10}$ with such a self-equivalence, which usually confines them to a much smaller linear space.
Up to conjugacy and power similarity, self-equivalences of prime order have 299 types.
We settle 298 of them.
Elementary lemmas exclude 280 types, and exhaustive searches confirm this for 242 of them.
The other 18 types are settled by exhaustive searches with exact equivalence certificates.
Up to extended affine equivalence, the 298 types contain 29 classes of quadratic APN functions, four of them new.
This also classifies the quadratic APN functions with coefficients in a subfield.
Exactly 24 classes contain one with coefficients in $\mathbb{F}_4$, and all 3 that contain one with coefficients in $\mathbb{F}_{32}$ are among them.
The open type consists of the quadratic canonical triplicate functions.
They correspond to maximum additive Hermitian rank-distance codes.
Three different searches find only the invariants of 16 classes, all among the 29.
We conjecture there are no others.
No quadratic APN function in the 298 settled types is Carlet--Charpin--Zinoviev equivalent to a permutation.
\end{abstract}


\section{Introduction}\label{sec:intro}

\IEEEPARstart{A}{} function $F\colon\F_2^n\to\F_2^n$ is \emph{almost perfect nonlinear} (APN) if the equation
$F(x+a)+F(x)=b$ has at most two solutions for every $a\neq 0$ and every $b$~\cite{DBLP:conf/eurocrypt/Nyberg93}.
So an APN function has the lowest possible differential uniformity.
For an S-box, this is the best possible property against differential cryptanalysis~\cite{DBLP:conf/eurocrypt/Nyberg93}.
APN permutations in odd dimension are used in practice, for example in the S-boxes of MISTY1~\cite{Matsui97}.
APN functions are also linked to many objects in coding theory and finite geometry~\cite{DBLP:books/cu/Carlet2021}.
Most known APN functions are \emph{quadratic}, that is, of algebraic degree two.

APN functions are classified up to CCZ-equivalence~\cite{DBLP:journals/dcc/CarletCZ98}.
For quadratic APN functions, CCZ-equivalence is the same as extended affine (EA) equivalence~\cite{DBLP:journals/jalc/Yoshiara12}.
In dimension $n\le 5$ all APN functions are known, and in dimension six all quadratic ones are known.
In dimension eight, searches have found millions of quadratic APN classes
\cite{DBLP:journals/tit/BeierleL22,DBLP:journals/ccds/IchanskaBKY26,DBLP:journals/corr/abs-2508-04644}.
Of the 3,808,491 classes listed in~\cite{DBLP:journals/corr/abs-2508-04644} (arXiv version~2), only 12,775 have an automorphism group larger than the translation group~\cite[Table~5.2]{DBLP:journals/corr/abs-2508-04644}.
By Section~\ref{sec:prelim}, the others have no nontrivial linear self-equivalence.
So at most 12,775 of them, that is, at most 0.34\%, have a nontrivial linear self-equivalence.
Dimension ten is different.
Only 25 classes of quadratic APN functions of $\F_2^{10}$ are known from published instances~\cite{DBLP:journals/ffa/BudaghyanCV20,DBLP:journals/tit/BeierleL22,DBLP:journals/ccds/BudaghyanIK23,HellesethIK26}, as collected in~\cite{Bellini26ccz}.
When we say ``published classes'', we mean these 25.
They all have algebraic structure: they are power functions, members of infinite families, or functions with a nontrivial self-equivalence.

Our starting point is an observation.
\emph{Each of the 25 known classes has a linear self-equivalence of prime order}
(Section~\ref{sec:results}).
A linear self-equivalence of $F$ is a pair of invertible linear maps $(A,B)$ such that $F\circ A$ and $B\circ F$ differ by an affine map.
Beierle and Leander~\cite{DBLP:journals/tit/BeierleL22} used such self-equivalences to find new quadratic APN functions in dimensions nine and ten.
They listed the possible types of prime-order self-equivalences and searched them, but not exhaustively.
In this paper we make their approach exhaustive in dimension ten.

\subsection{Why dimension ten}
The \emph{big APN problem} asks whether there is an APN permutation of $\F_2^n$ for some even $n\ge 8$.
The only known example in even dimension lives in dimension six~\cite{BrowningDMW10}.
It is CCZ-equivalent to a quadratic APN function with many self-equivalences.
We ask whether a similar quadratic function exists in dimension eight or ten.
Table~\ref{tab:n8n10} compares the two dimensions.
In dimension eight, millions of quadratic APN classes are known.
None of them is CCZ-equivalent to a permutation~\cite{DBLP:journals/tit/BeierleL22,DBLP:journals/corr/abs-2508-04644}.
In dimension ten, randomized searches have found few classes~\cite{DBLP:journals/tit/BeierleL22}.
The known classes are structured, and all of them have a nontrivial self-equivalence.
So we classify the quadratic APN functions of dimension ten by their self-equivalences.
Dimension ten is also the next case after six with $n\equiv 2\pmod 4$.

\begin{table}[t]
\centering
\caption{Quadratic APN functions in dimensions eight and ten: two opposite situations.}\label{tab:n8n10}
\begin{tabular}{@{}p{0.36\linewidth}p{0.27\linewidth}p{0.27\linewidth}@{}}
\toprule
 & $n=8$ & $n=10$ \\
\midrule
Known quadratic APN classes & millions & 25 (29 with this paper) \\
Random search finds new classes & yes & rarely \\
Classes with a self-equivalence of prime order & at most 12,775 of 3,808,491 (0.34\%) in~\cite{DBLP:journals/corr/abs-2508-04644} & all known ones \\
Where the known classes come from & large computational searches and constructions & power functions, infinite families, self-equivalences \\
Classes CCZ-equivalent to a permutation & none~\cite{DBLP:journals/tit/BeierleL22,DBLP:journals/corr/abs-2508-04644} & none; three pass the counting criterion but fail the projection test~\cite{Bellini26ccz} \\
Analogy with Dillon's permutation ($n=6$) & $n\equiv 0 \pmod 4$ & $n\equiv 2\pmod 4$, as $n=6$ \\
\bottomrule
\end{tabular}
\end{table}

\subsection{Contributions}
Let $\beta(x,y)=F(x+y)+F(x)+F(y)+F(0)$ be the polar form of a quadratic function $F$.
A pair $(A,B)$ of invertible linear maps is a self-equivalence of $F$ if and only if $\beta(Ax,Ay)=B\,\beta(x,y)$ for all $x,y$.
Every nontrivial self-equivalence has a power of prime order.
Up to conjugacy and power similarity there are 299 types of prime-order pairs $(A,B)$ in dimension ten
(Section~\ref{sec:types}).
Our main result is the following.

\begin{theorem}[computer-assisted]\label{thm:main}
Let $F$ be a quadratic APN function of $\F_2^{10}$ with a linear self-equivalence of prime order.
Suppose that $F$ has such a self-equivalence of a type other than type~64.
Then $F$ is EA-equivalent to one of 29 functions.
They are the 25 published classes and four new ones.
Each of the 29 classes occurs in at least one type other than type~64.
\end{theorem}

\begin{figure}[t]
\centering
\begin{tikzpicture}[
  box/.style={draw, rounded corners=2pt, align=center, inner sep=4pt, font=\small, minimum height=1.1cm},
  arr/.style={-{Stealth[length=2mm]}, thick}]
\node[box, fill=gray!12, text width=2.6cm] (all) at (0,0) {299 types of\\ self-equivalence\\ of prime order};
\node[box, fill=blue!10, text width=3.4cm] (proof) at (4.6,1.7) {\textbf{280 types}\\ excluded by proof\\ (lemmas on the polar form)};
\node[box, fill=orange!14, text width=3.4cm] (search) at (4.6,0) {\textbf{18 types}\\ exhaustive search\\ ($\approx$ 36,400 core-hours)};
\node[box, fill=red!10, text width=3.4cm] (open) at (4.6,-1.7) {\textbf{1 type} (64)\\ canonical triplicates\\ = Hermitian codes};
\node[box, fill=green!12, text width=3.3cm] (res) at (9.4,0.85) {\textbf{29 classes}\\ 25 published + 4 new\\ none CCZ to a permutation};
\node[box, fill=green!6, text width=3.3cm] (conj) at (9.4,-1.7) {16 known classes;\\ no new fingerprint found\\ (conjecture: no others)};
\draw[arr] (all) -- (proof.west);
\draw[arr] (all) -- (search.west);
\draw[arr] (all) -- (open.west);
\draw[arr] (search.east) -- (res.west);
\draw[arr] (open.east) -- (conj.west);
\node[font=\small, gray] at (4.6,2.85) {no APN function};
\end{tikzpicture}
\caption{Overview of the classification. The core-hours cover the classification and confirmation searches, including the confirmation searches on 242 types that are also excluded by proof (Section~\ref{sec:search}).}\label{fig:overview}
\end{figure}
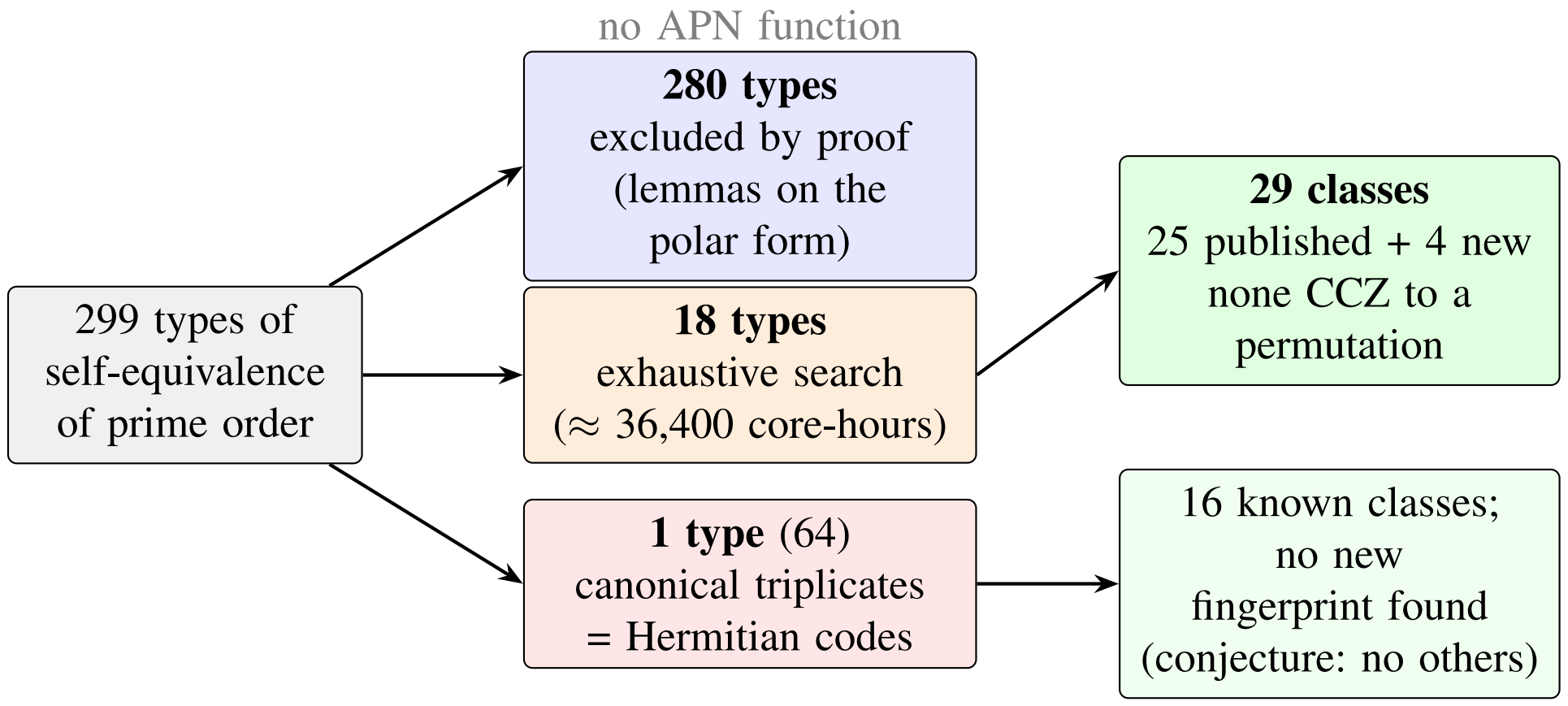

Figure~\ref{fig:overview} gives an overview of the classification.
\begin{itemize}
\item \emph{Proofs.}
We give elementary lemmas on the polar form (Section~\ref{sec:lemmas}).
They show that 280 types contain no APN function.
A bound of Baudrin et al.~\cite{BaudrinBFLNPS25} gives a partial cross-check.
\item \emph{Exhaustive searches.}
The other 18 types are settled by exhaustive searches (Sections~\ref{sec:search} and~\ref{sec:special}).
Two of them, types~24 and~47, need a dedicated method.
We also ran the generic search exhaustively on 242 of the 280 excluded types, and it agrees with the proofs.
These searches are validated on cases with a known answer.
Every function they find is matched to a known class by an explicit equivalence certificate.
\item \emph{One open type.}
Type~64 consists of the quadratic \emph{canonical triplicate} functions~\cite{DBLP:journals/ccds/BudaghyanIK23}.
We show that they correspond to maximum additive Hermitian rank-distance codes (Section~\ref{sec:type64}).
Three exploratory searches find only functions with the fingerprint, an EA-invariant, of one of 16 classes, all among the 29.
Unlike the searches that settle the other 298 types, these three searches are not exhaustive on type~64, and they do not certify equivalences.
We conjecture that there are no others.
If the conjecture holds, the quadratic APN functions of $\F_2^{10}$ with a nontrivial linear self-equivalence are exactly the 29 classes.
\end{itemize}

Two special cases are of independent interest.
Type~73 is the set of quadratic functions of $\F_{2^{10}}$ with coefficients in $\F_4$.
Ichanska et al.\ considered an exhaustive search of this set infeasible~\cite{DBLP:journals/ccds/IchanskaBKY26}.
We find exactly 24 APN classes in it, four of them new (Theorem~\ref{thm:t73}).
Type~24 is the set of quadratic functions with coefficients in $\F_{32}$.
In dimension six, the analogous set contains a function CCZ-equivalent to Dillon's permutation.
It contains only three APN classes (Theorem~\ref{thm:t24}).
The proper subfields of $\F_{2^{10}}$ are $\F_2$, $\F_4$ and $\F_{32}$.
So these two cases classify the quadratic APN functions of $\F_{2^{10}}$ with coefficients in any proper subfield.
For $\F_2$ we find the three classes of Yu et al.~\cite{YuLIK25}.

None of the 29 classes is CCZ-equivalent to a permutation.
For the 25 published classes this was known~\cite{DBLP:journals/tit/BeierleL22,DBLP:journals/ccds/BudaghyanIK23,Bellini26ccz}, and we check it for the four new ones.
So, within the reach of this classification, there is no quadratic route to an APN permutation of $\F_2^{10}$.

\subsection{Related work}
Beierle and Leander introduced the search by self-equivalence types~\cite{DBLP:journals/tit/BeierleL22}.
Baudrin, Canteaut and Perrin describe the linear self-equivalences of the known APN families~\cite{BaudrinCP26}.
Baudrin et al.\ prove a general bound on the fixed spaces of a self-equivalence of a differentially uniform function~\cite{BaudrinBFLNPS25}.
Beierle, Brinkmann and Leander study the self-equivalences of APN permutations~\cite{DBLP:journals/tit/BeierleBL21}.
Göloğlu and Kölsch determine the equivalences between biprojective APN functions~\cite{DBLP:journals/combtheory/GologluK25}.
Dempwolff and Edel study the automorphisms of extensions of quadratic APN functions and of dimensional dual hyperovals.
Their methods rely on elementary abelian 2-subgroups of the automorphism group~\cite{DempwolffE16}.
Budaghyan, Ivkovic and Kaleyski introduced triplicate functions~\cite{DBLP:journals/ccds/BudaghyanIK23}.
Ichanska et al.\ search quadratic functions with coefficients in subfields~\cite{DBLP:journals/ccds/IchanskaBKY26}.
Yu et al.\ classify the quadratic APN functions of $\F_{2^{10}}$ with coefficients in $\F_2$~\cite{YuLIK25}.
Helleseth, Ichanska and Kaleyski give new infinite families in even dimension~\cite{HellesethIK26}.
Our tools for equivalence follow Canteaut, Couvreur and Perrin~\cite{DBLP:journals/tit/CanteautCP22}.

\subsection{Organization}
Section~\ref{sec:prelim} recalls the background.
Section~\ref{sec:types} lists the 299 types.
Section~\ref{sec:lemmas} gives the lemmas that exclude types by proof.
Section~\ref{sec:search} describes the generic exhaustive search and the equivalence test.
Section~\ref{sec:special} treats the types that need a dedicated method.
Section~\ref{sec:type64} studies the open type~64.
Section~\ref{sec:results} collects the results, and Section~\ref{sec:conclusion} concludes.

\section{Preliminaries}\label{sec:prelim}

\subsection{Quadratic APN functions}
We identify $\F_2^n$ with the field $\F_{2^n}$ when convenient.
Write $F\colon\F_2^n\to\F_2^n$ as $F=(f_1,\dots,f_n)$ with coordinate functions $f_i\colon\F_2^n\to\F_2$.
The \emph{algebraic degree} of $F$ is the maximum degree of the algebraic normal forms of $f_1,\dots,f_n$.
In univariate form $F(x)=\sum_i a_i x^i$ over $\F_{2^n}$, it is the maximum binary weight of an exponent $i$ with $a_i\neq 0$.
The function $F$ is \emph{quadratic} if all its derivatives
$D_aF(x)=F(x+a)+F(x)$ are affine, that is, if its algebraic degree is at most two.
For $n\ge 2$ an affine function is not APN, so a quadratic APN function has algebraic degree exactly two.
The \emph{polar form} of a quadratic $F$ is
\[
\beta(x,y)=F(x+y)+F(x)+F(y)+F(0).
\]
It is bilinear, symmetric and alternating: $\beta(x,x)=0$.
Two quadratic functions with the same polar form differ by an affine map.
So, up to EA-equivalence, a quadratic function is given by its polar form.
We write $\beta_a=\beta(a,\cdot)$ for the linear map $y\mapsto\beta(a,y)$.
Since $D_aF(x)=\beta_a(x)+D_aF(0)$, the following criterion is standard.

\begin{lemma}\label{lem:apn-kernel}
A quadratic function $F$ is APN if and only if $\ker\beta_a=\{0,a\}$ for every $a\neq 0$.
Equivalently, $\beta(x,y)\neq 0$ for all linearly independent $x,y$.
\end{lemma}

For a linear form $c\in(\F_2^n)^*$, the \emph{component} $c\cdot F$ is a quadratic Boolean function.
Its polar form $c\cdot\beta$ is an alternating bilinear form.
The component is bent if and only if $c\cdot\beta$ is nondegenerate.

\subsection{Equivalences and invariants}
Two functions $F,G$ are \emph{EA-equivalent} if $G=L_1\circ F\circ L_2+L_3$ with $L_1,L_2$ affine permutations and $L_3$ affine.
EA-equivalence preserves the algebraic degree of functions of degree at least two.
They are \emph{CCZ-equivalent} if an affine permutation of $\F_2^{2n}$ maps the graph of $F$ onto the graph of $G$~\cite{DBLP:journals/dcc/CarletCZ98}.
For quadratic APN functions the two notions agree~\cite{DBLP:journals/jalc/Yoshiara12}.
We use the invariants of Canteaut, Couvreur and Perrin~\cite{DBLP:journals/tit/CanteautCP22}.
The \emph{ortho-derivative} of a quadratic APN function maps $a\neq 0$ to the unique nonzero linear form $\pi(a)$ that vanishes on the image of $\beta_a$.
The differential and extended Walsh spectra of $\pi$ are EA-invariants of $F$.
In this paper, a \emph{class} is an EA-class of quadratic APN functions.
By~\cite{DBLP:journals/jalc/Yoshiara12}, it is exactly the set of quadratic functions in a CCZ-class.
That CCZ-class may also contain functions of higher degree.
For example, Dillon's APN permutation of $\F_2^6$ has degree four, and its CCZ-class contains a quadratic APN function~\cite{BrowningDMW10}.
So the number of classes is both the number of EA-classes of quadratic APN functions and the number of CCZ-classes that contain a quadratic APN function.
All counts of classes in this paper, including the counts of known classes we cite, are of this kind.
APN functions whose CCZ-class contains no quadratic function are not counted.
To show that two functions are in the same class, we always compute an explicit equivalence and check it.
To show that they are in different classes, we compare invariants.

\subsection{Permutations}
A component $c\cdot F$ is \emph{non-bent} if $c\cdot\beta$ is degenerate.
Let $\mathcal{N}$ be the set of linear forms $c$ that give non-bent components, together with $0$.
If $F$ is CCZ-equivalent to a permutation, then $\mathcal{N}$ contains two linear subspaces $B_1,B_2$ with $B_1+B_2=(\F_2^n)^*$~\cite{DBLP:journals/ccds/GologluP21}.
We call this the \emph{counting criterion}.
It implies a weaker condition: the largest dimension $M$ of a linear subspace contained in $\mathcal{N}$ satisfies $M\ge n/2$.
When we only use this bound, we say so.
In~\cite{Bellini26ccz} we give an exact test, the \emph{projection test}, and apply both tests to the 25 published classes in dimension ten collected there.
We use the same tests for the new classes of this paper.

\subsection{Self-equivalences}
A \emph{linear self-equivalence} of a quadratic function $F$ is a pair $(A,B)$ of invertible linear maps of $\F_2^n$ such that $F\circ A+B\circ F$ is affine.
Equivalently,
\begin{equation}\label{eq:equivariant}
\beta(Ax,Ay)=B\,\beta(x,y)\qquad\text{for all }x,y.
\end{equation}
The pairs that satisfy~\eqref{eq:equivariant} form a group.
If $(A,B)$ is a self-equivalence of $F$, then $(TAT^{-1},SBS^{-1})$ is a self-equivalence of the function with polar form $S\beta(T^{-1}x,T^{-1}y)$,
which is EA-equivalent to $F$.
So only the conjugacy classes of $A$ and $B$ matter.
Every nontrivial element of a finite group has a power of prime order.
So every quadratic function with a nontrivial linear self-equivalence has one of prime order $p$,
that is, $A^p=I$, $B^p=I$ and $(A,B)\neq(I,I)$.
In this paper ``self-equivalence'' always means a linear one.

Other papers count the full automorphism group $\mathrm{Aut}(F)$, the group of affine permutations of $\F_2^{2n}$ that preserve the graph of $F$~\cite{DBLP:journals/corr/abs-2508-04644}.
For a quadratic $F$, it contains the maps $(x,y)\mapsto(x+a,\,y+D_aF(x))$, which are affine since $D_aF$ is affine.
They form a group of order $2^n$.
If $F(Ax)=BF(x)+\ell(x)$ with $\ell$ affine, then $(x,y)\mapsto(Ax,\,By+\ell(x))$ is also in $\mathrm{Aut}(F)$.
Composing it with the $2^n$ maps above gives $2^n$ elements with linear parts $A$ and $B$.
So $|\mathrm{Aut}(F)|\ge 2^n\,|G|$, where $G$ is the group of pairs $(A,B)$ that satisfy~\eqref{eq:equivariant}.
In particular, if $|\mathrm{Aut}(F)|=2^n$, then $F$ has no nontrivial linear self-equivalence.

For fixed $(A,B)$, the polar forms that satisfy~\eqref{eq:equivariant} form a linear space
\[
W(A,B)=\{\beta \text{ alternating bilinear} : \beta(Ax,Ay)=B\,\beta(x,y)\}.
\]
It is the solution space of a homogeneous linear system with $n\binom{n}{2}$ unknowns, the coordinates of the vectors $\beta(e_i,e_j)$, $i<j$.
For $n=10$ there are 450 unknowns.
Every search and every proof in this paper starts from $W(A,B)$.

\subsection{Notation}
We write $\Fix(A)=\ker(A+I)$ for the fixed space of $A$.
For a polynomial $g$ we write $\ker g(A)$ for the kernel of $g$ evaluated at $A$.
Let $A$ have odd prime order $p$.
Then $\F_2^n$ is the direct sum of $\Fix(A)$ and the spaces $\ker\varphi(A)$.
Here $\varphi$ runs over the irreducible factors of the cyclotomic polynomial $\Phi_p$.
We call $\ker\varphi(A)$ the \emph{$\varphi$-part} of $A$.
Each factor has degree $\mathrm{ord}_p(2)$, the multiplicative order of 2 modulo $p$.
We write $\omega$ for a primitive element of $\F_4$.
We write $\Tr$ for the absolute trace from $\F_{2^n}$ to $\F_2$.

\section{The 299 types}\label{sec:types}

We follow Beierle and Leander~\cite[Sec.~4.1]{DBLP:journals/tit/BeierleL22}.
Let $p$ be a prime.
An element of order $p$ of $\GL(10,2)$ is described, up to conjugacy, as follows.
\begin{itemize}
\item If $p$ is odd, it is a block-diagonal matrix.
Each block is the companion matrix of an irreducible factor of $\Phi_p$.
An identity block completes the matrix.
So the conjugacy class is given by a multiset of factors of $\Phi_p$ of total degree at most 10.
All factors have degree $d=\mathrm{ord}_p(2)$, so we need $d\le 10$.
This leaves $p\in\{3,5,7,11,17,31,73,127\}$, with $d=2,4,3,10,8,5,9,7$.
\item If $p=2$, it is an involution.
Its conjugacy class is given by the number $k\in\{1,\dots,5\}$ of Jordan blocks of size two.
We write $J_k$ for it. Then $\dim\Fix(J_k)=10-k$.
\end{itemize}
A self-equivalence $(A,B)$ of prime order has one of three kinds:
\begin{itemize}
\item $(p,p)$, if both $A$ and $B$ have order $p$;
\item $(p,1)$, if $B=I$;
\item $(1,p)$, if $A=I$.
\end{itemize}
Conjugating $A$ and $B$ independently does not change the EA-class of the functions (Section~\ref{sec:prelim}).
Moreover, $(A,B)$ and $(A^r,B^r)$ with $r$ coprime to $p$ generate the same group.
So we also identify pairs up to this \emph{power similarity}.
It permutes the factors of $\Phi_p$: the factor with root $z$ goes to the factor with root $z^r$.

\begin{table}[t]
\centering
\caption{Number of self-equivalence types of prime order in dimension ten.}\label{tab:types}
\begin{tabular}{@{}rrrrrr@{}}
\toprule
$p$ & $\mathrm{ord}_p(2)$ & $(p,p)$ & $(p,1)$ & $(1,p)$ & total \\
\midrule
2   & --  & 25  & 5  & 5  & 35 \\
3   & 2   & 25  & 5  & 5  & 35 \\
5   & 4   & 4   & 2  & 2  & 8 \\
7   & 3   & 41  & 5  & 5  & 51 \\
11  & 10  & 1   & 1  & 1  & 3 \\
17  & 8   & 2   & 1  & 1  & 4 \\
31  & 5   & 123 & 5  & 5  & 133 \\
73  & 9   & 8   & 1  & 1  & 10 \\
127 & 7   & 18  & 1  & 1  & 20 \\
\midrule
all &     & 247 & 26 & 26 & 299 \\
\bottomrule
\end{tabular}
\end{table}

Table~\ref{tab:types} gives the number of types.
We obtained it by direct enumeration.
For each $p$, we list the conjugacy classes as above, form all pairs, and count the orbits of the powers $r\in(\mathbb{Z}/p)^*$.
The power $r$ maps a factor of $\Phi_p$ with root $z$ to the factor with root $z^r$.
A short independent script, included in the artifact, reproduces every entry of the table.
We number them from 0 to 298, in the order of the artifact~\cite{artifact}.
The 247 types of kind $(p,p)$ agree with the list of Beierle and Leander~\cite{DBLP:journals/tit/BeierleL22}.
Their list of one-sided types is not reduced by power similarity, so it is longer than ours.

\subsection{The space of a type}
For each type we fix matrices $A$ and $B$ and compute the space $W(A,B)$ of~\eqref{eq:equivariant}.
Its dimension ranges from 0 to 405.
Thirty-one types have $W(A,B)=0$: they contain no nonzero polar form, hence no quadratic function of degree two.
For the others, the size of an exhaustive search depends on much more than $\dim W(A,B)$.
We estimate it with Knuth's random probing method~\cite{Knuth75} on the search tree of Section~\ref{sec:search}.
The estimates range from a single node to about $6\cdot 10^{72}$ nodes.
So a search alone cannot settle all types, and we first exclude as many types as possible by proof.

\begin{table}[t]
\centering
\caption{The 19 types not excluded by proof. For odd $p$, a block structure lists the factors of $\Phi_p$ with multiplicity; an identity block completes the matrix. $J_k$ is an involution with $k$ Jordan blocks of size two, and $I$ the identity.}\label{tab:keytypes}
\begin{tabular}{@{}rrll@{\qquad}rrll@{}}
\toprule
Type & $p$ & $A$ & $B$ & Type & $p$ & $A$ & $B$ \\
\midrule
24  & 2  & $J_5$ & $J_5$ & 175 & 31 & $\varphi_1^2$ & $\varphi_3^2$ \\
47  & 3  & $\Phi_3^3$ & $\Phi_3^3$ & 176 & 31 & $\varphi_1^2$ & $\varphi_3\varphi_5$ \\
64  & 3  & $\Phi_3^5$ & $I$ & 180 & 31 & $\varphi_1^2$ & $\varphi_5^2$ \\
73  & 5  & $\Phi_5^2$ & $\Phi_5^2$ & 198 & 31 & $\varphi_1\varphi_3$ & $\varphi_1\varphi_5$ \\
116 & 7  & $\psi_1^2\psi_3$ & $\psi_1^2\psi_3$ & 207 & 31 & $\varphi_1\varphi_3$ & $\varphi_5^2$ \\
129 & 11 & $\Phi_{11}$ & $\Phi_{11}$ & 208 & 31 & $\varphi_1\varphi_3$ & $\varphi_5\varphi_7$ \\
132 & 17 & $\chi$ & $\chi$ & 209 & 31 & $\varphi_1\varphi_3$ & $\varphi_5\varphi_{11}$ \\
170 & 31 & $\varphi_1^2$ & $\varphi_1\varphi_3$ & 231 & 31 & $\varphi_1\varphi_5$ & $\varphi_3\varphi_7$ \\
171 & 31 & $\varphi_1^2$ & $\varphi_1\varphi_5$ & 235 & 31 & $\varphi_1\varphi_5$ & $\varphi_5\varphi_7$ \\
    &    &               &                       & 255 & 31 & $\varphi_1\varphi_{15}$ & $\varphi_3\varphi_7$ \\
\bottomrule
\end{tabular}
\end{table}

Table~\ref{tab:keytypes} gives the block structure of the types that are not excluded by proof, so that the numbering can be checked without the artifact.
For $p=7$, $\psi_1=x^3+x+1$ and $\psi_3=x^3+x^2+1$.
For $p=17$, $\chi$ is one of the two factors of degree eight of $\Phi_{17}$; power similarity exchanges them.
For $p=31$, let $\zeta$ be a root of $x^5+x^2+1$; then $\varphi_r$ is the minimal polynomial of $\zeta^r$:
$\varphi_1=x^5+x^2+1$, $\varphi_3=x^5+x^4+x^3+x^2+1$, $\varphi_5=x^5+x^4+x^2+x+1$, $\varphi_7=x^5+x^3+x^2+x+1$, $\varphi_{11}=x^5+x^4+x^3+x+1$ and $\varphi_{15}=x^5+x^3+1$.
The artifact gives the matrices of all 299 types~\cite{artifact}.

\subsection{Some types in plain words}
Several types have a simple description in the field $\F_{2^{10}}$.
We use the following ones below.
\begin{itemize}
\item \emph{Type~24} is $(J_5,J_5)$.
$J_5$ is conjugate to the Frobenius map $x\mapsto x^{32}$.
By Lemma~\ref{lem:frob} below, type~24 is the set of quadratic functions with coefficients in $\F_{32}$, up to EA-equivalence.
\item \emph{Type~73} is $(A,A)$ with $A$ of order 5, with two blocks of $\Phi_5$ and $\dim\Fix(A)=2$.
$A$ is conjugate to the Frobenius map $x\mapsto x^4$.
By Lemma~\ref{lem:frob}, type~73 is the set of quadratic functions with coefficients in $\F_4$, up to EA-equivalence.
\item \emph{Type~64} is $(A,I)$ with $A$ the multiplication by $\omega$ on $\F_4^5$.
These are the functions with $F(\omega x)=F(x)$ up to affine terms.
They are the quadratic canonical triplicate functions~\cite{DBLP:journals/ccds/BudaghyanIK23,KolschKK23,BaudrinCP26}.
\item \emph{Types~41 and~47} are $(A,A)$ with $A$ of order 3, with two and three blocks of $\Phi_3$.
So $\dim\Fix(A)$ is 6 and 4.
\end{itemize}

Types 24 and 73 hold the self-equivalence only up to affine terms.
The next lemma shows that this makes no difference for subfield coefficients.

\begin{lemma}\label{lem:frob}
Let $k$ divide $n$, let $\phi(x)=x^{2^k}$ on $\F_{2^n}$, and let $F$ be a quadratic function of $\F_{2^n}$.
Write $F=Q+L$, where $Q$ collects the terms of the reduced univariate polynomial of $F$ whose exponents have binary weight two, and $L$ is affine.
Then $(\phi,\phi)$ is a self-equivalence of $F$ if and only if all coefficients of $Q$ lie in $\F_{2^k}$.
In that case $F$ is EA-equivalent to $Q$, which satisfies $Q\circ\phi=\phi\circ Q$ exactly.
\end{lemma}

\begin{proof}
Write $Q(x)=\sum_e c_ex^e$ over the exponents $e$ of binary weight two.
Then $Q(\phi(x))+\phi(Q(x))=\sum_e(c_e+c_e^{2^k})x^{2^ke}$, where $2^ke$ is reduced modulo $2^n-1$.
The map $e\mapsto 2^ke$ permutes the exponents of weight two.
So this is a reduced polynomial whose exponents all have weight two.
It is affine only if it is zero, that is, only if $c_e\in\F_{2^k}$ for every $e$.
Since $L$ is affine, $F\circ\phi+\phi\circ F$ is affine if and only if $Q\circ\phi+\phi\circ Q$ is.
\end{proof}

A function has a self-equivalence of type~24 if and only if it is EA-equivalent to one with the self-equivalence $(\phi,\phi)$ for $k=5$, since $\phi$ is conjugate to $J_5$.
By Lemma~\ref{lem:frob}, these are the quadratic functions with coefficients in $\F_{32}$, up to EA-equivalence.
The same holds for type~73 with $k=2$ and $\F_4$.

\section{Excluding types by proof}\label{sec:lemmas}

In this section $F$ is a quadratic APN function of $\F_2^n$ with polar form $\beta$,
and $(A,B)$ is a self-equivalence of prime order, so~\eqref{eq:equivariant} holds.
The lemmas give necessary conditions on the type of $(A,B)$.
Each condition can be checked on the matrices $A$, $B$ or on the space $W(A,B)$.
When a condition fails, the type contains no APN function, and no search is needed.

\subsection{Intertwiners}
The first lemma uses the fixed points of $A$.

\begin{lemma}[Intertwiner lemma]\label{lem:intertwiner}
Let $x\in\Fix(A)$, $x\neq 0$, and let $T_x=\beta_x$.
Then $T_xA=BT_x$ and $\ker T_x=\{0,x\}$.
Consequently:
\begin{enumerate}
\item[(i)] For every polynomial $g$, $T_x$ maps $\ker g(A)$ into $\ker g(B)$, and
$\dim\ker g(A)-[x\in\ker g(A)]\le\dim\ker g(B)$.
Here $[x\in V]$ is $1$ if $x\in V$ and $0$ otherwise.
\item[(ii)] If $p$ is odd and $\Fix(A)\neq 0$, then $\dim\ker\varphi(A)\le\dim\ker\varphi(B)$
for every irreducible factor $\varphi$ of $\Phi_p$.
\item[(iii)] If $A=J_{k_A}$ and $B=J_{k_B}$ are involutions, then $k_A\le k_B$ if $k_A<n/2$, and $k_A-1\le k_B$ if $k_A=n/2$.
\end{enumerate}
\end{lemma}

\begin{proof}
For all $y$ we have $T_x(Ay)=\beta(Ax,Ay)=B\beta(x,y)=BT_x(y)$, since $Ax=x$.
The kernel of $T_x$ is $\{0,x\}$ by Lemma~\ref{lem:apn-kernel}.
From $T_xA=BT_x$ we get $T_x\,g(A)=g(B)\,T_x$, so $T_x$ maps $\ker g(A)$ into $\ker g(B)$.
Its kernel on $\ker g(A)$ is $\{0,x\}\cap\ker g(A)$, which gives (i).
For (ii), take $g=\varphi$: then $x\notin\ker\varphi(A)$, because $x$ is fixed and $\varphi(1)\neq 0$.
For (iii), put $N_A=A+I$ and $N_B=B+I$.
Then $T_xN_A=N_BT_x$, so $T_x$ maps $\Img N_A$ into $\Img N_B$, with kernel $\{0,x\}\cap\Img N_A$.
We have $\dim\Img N_A=k_A$ and $\Img N_A\subseteq\Fix(A)$.
If $k_A<n/2$, then $\Img N_A\neq\Fix(A)$, and we may choose $x\in\Fix(A)\setminus\Img N_A$.
This gives $k_A\le k_B$. If $k_A=n/2$, every choice gives $k_A-1\le k_B$.
\end{proof}

Lemma~\ref{lem:intertwiner} says that the block structure of $A$ must fit into that of $B$.
It does not use the space $W(A,B)$ at all.
The next lemma looks at the action of $\beta$ in the other direction.
It fixes a point outside $\Fix(A)$ and reads $\beta$ on $\Fix(A)$.

\begin{lemma}[Dual intertwiner lemma]\label{lem:dual}
Let $p$ be odd, and let $\varphi$ be an irreducible factor of $\Phi_p$ with $\ker\varphi(A)\neq 0$.
Then $\dim\Fix(A)\le\dim\ker\varphi(B)$.
\end{lemma}

\begin{proof}
For $z\in\Fix(A)$ and all $u$ we have $\beta(Au,z)=\beta(Au,Az)=B\beta(u,z)$.
So $u\mapsto\beta(u,z)$ intertwines $A$ and $B$, and it maps $\ker\varphi(A)$ into $\ker\varphi(B)$.
Fix $u\in\ker\varphi(A)$, $u\neq 0$.
The map $z\mapsto\beta(u,z)$ is linear on $\Fix(A)$, with values in $\ker\varphi(B)$.
Its kernel is contained in $\{0,u\}\cap\Fix(A)=\{0\}$, since $u$ is not fixed.
So it is injective.
\end{proof}

Baudrin et al.\ proved a bound of a similar kind for all functions of low differential uniformity,
not only quadratic ones~\cite[Corollary~3 and Remark~1]{BaudrinBFLNPS25}.
Their self-equivalences satisfy $F\circ A=B\circ F$ exactly.
For an APN function with such a nontrivial self-equivalence, the bound reads
\begin{equation}\label{eq:bbf}
\dim\Fix(A)+\dim\Fix(B)\le n+1.
\end{equation}
Our self-equivalences hold only up to an affine term.
So~\eqref{eq:bbf} does not apply to our types directly, and we only compare it with our necessary conditions.
The two bounds are different: \eqref{eq:bbf} compares $\Fix(A)$ with $\Fix(B)$,
while Lemma~\ref{lem:dual} compares $\Fix(A)$ with one part of $B$.
In dimension ten, Lemma~\ref{lem:dual} excludes 95 types, and~\eqref{eq:bbf} would exclude 65.
Each excludes types that the other does not.
For instance, type~41 has $\dim\Fix(A)=\dim\Fix(B)=6$, so it violates~\eqref{eq:bbf}.
It is also excluded by Lemma~\ref{lem:dual}, since $\dim\ker\varphi(B)=4<6$.
Neither bound excludes a type beyond Table~\ref{tab:proofs} below.

\subsection{Restriction to an invariant subspace}
The values of $\beta$ on fixed vectors are fixed:
for $x,y\in\Fix(A)$ we have $B\beta(x,y)=\beta(Ax,Ay)=\beta(x,y)$.
So $F$ restricts to a quadratic APN function from $\Fix(A)$ to $\Fix(B)$, up to affine terms.
The next lemma bounds such restrictions.

\begin{lemma}\label{lem:restriction}
Let $k\ge 3$, and let $D\subseteq\F_2^n$ be a subspace of dimension $k$.
Let $E$ be the span of $\beta(D,D)$.
If $F$ is APN, then $\dim E\neq k-1$.
\end{lemma}

\begin{proof}
For $a\in D$, $a\neq 0$, the restriction of $\beta_a$ to $D$ has kernel $\{0,a\}$, so its image in $E$ has dimension $k-1$.
Suppose $\dim E=k-1$. Then this image is all of $E$ for every $a$.
Take a nonzero linear form $e^*$ on $E$.
The alternating form $e^*\circ\beta$ on $D$ is nondegenerate.
Indeed, if $a\neq 0$ is in its radical, then $e^*$ vanishes on $\beta_a(D)=E$, so $e^*=0$.
So the forms $e^*\circ\beta$, $e^*\in E^*$, make up a space of dimension $k-1$ of alternating forms on $\F_2^k$ whose nonzero elements are all nondegenerate.
If $k$ is odd, this is impossible, since every alternating form in odd dimension is degenerate.
If $k$ is even, the Pfaffian is a form of degree $k/2$ in the $k-1$ coordinates of $e^*$.
Since $k/2<k-1$, Chevalley's theorem~\cite[Cor.~6.6]{LidlN97} gives a nonzero $e^*$ with Pfaffian zero, that is, a degenerate form.
\end{proof}

Take $D=\Fix(A)$, so that $E\subseteq\Fix(B)$.
The proof shows that $E$ contains $\beta_a(D)$, of dimension $\dim D-1$.
So $\dim\Fix(B)=\dim\Fix(A)-1$ forces $E=\Fix(B)$, which Lemma~\ref{lem:restriction} forbids when $\dim\Fix(A)\ge 3$.
Together with the obvious bound $\dim\Fix(B)\ge\dim\Fix(A)-1$, this extends Lemma~4 of Beierle and Leander~\cite{DBLP:journals/tit/BeierleL22},
which states $\dim\Fix(B)\ge\dim\Fix(A)$ outside two small cases.
The bound on spaces of nondegenerate alternating forms is also noted by Gow~\cite[Sec.~3]{Gow17}.
In the computer check, $D$ runs over the sums of the subspaces $\ker(g(A)^j)$, where $g$ is an irreducible factor of $x^p+1$ and $j\in\{1,2\}$.
These subspaces are invariant under $A$.
The computer check does not use one candidate $\beta$ but the whole space $W(A,B)$.
It computes $E_W(D)=\spn\{\beta(d,e):\beta\in W(A,B),\ d,e\in D\}$ from a basis of $W(A,B)$.
For every $\beta\in W(A,B)$ the space $E$ of Lemma~\ref{lem:restriction} lies in $E_W(D)$, and it has dimension at least $k-1$ if $F$ is APN.
So $\dim E_W(D)\le k-1$ forces $\dim E=k-1$, which the lemma forbids.

\subsection{Order three}
For $p=3$ we have $\Phi_3=x^2+x+1$, and $U=\ker(A^2+A+I)$ is the part of $A$ without fixed points.

\begin{lemma}\label{lem:order3}
Let $(A,B)$ have order $p=3$, so that $A^3=B^3=I$, and let $U=\ker(A^2+A+I)$. If $F$ is APN, then $2\dim\Fix(B)\ge\dim U$.
\end{lemma}

\begin{proof}
For $x\in U$ put $q(x)=\beta(x,Ax)$.
Since $A^2x=Ax+x$, we get $Bq(x)=\beta(Ax,A^2x)=\beta(Ax,Ax+x)=\beta(Ax,x)=q(x)$.
So $q$ maps $U$ into $\Fix(B)$.
If $x\neq 0$, then $x$ and $Ax$ are linearly independent, so $q(x)\neq 0$ by Lemma~\ref{lem:apn-kernel}.
So the $\dim\Fix(B)$ coordinates of $q$ are quadratic forms in $\dim U$ variables with only the trivial common zero.
By the Chevalley--Warning theorem, $2\dim\Fix(B)\ge\dim U$.
\end{proof}

\subsection{Odd dimension over $\F_4$}

\begin{lemma}\label{lem:odd}
Let $n=2m$ with $m$ odd, and let $A$ be the multiplication by $\omega$ on $\F_4^m$.
Let $(A,B)$ have prime order; since $A$ has order 3, this order is 3, so $B^3=I$.
If $B\neq I$, then no quadratic APN function has the self-equivalence $(A,B)$.
\end{lemma}

\begin{proof}
Write the output space as $\Fix(B)\oplus Y$ with $Y=\ker(B^2+B+I)$.
Then $Y$ is an $\F_4$-space of dimension $r\ge 1$, where $B$ acts as $\omega$, and $\dim\Fix(B)=2m-2r$.
Let $\beta_\omega$ be the $Y$-component of $\beta$. It satisfies $\beta_\omega(\omega x,\omega y)=\omega \beta_\omega(x,y)$.
Extend scalars from $\F_2$ to $\F_4$, that is, work in $\F_4\otimes_{\F_2}\F_2^{n}$.
There the $\F_2$-linear map $A$ splits the space into its eigenspaces for $\omega$ and $\omega^2$, and so does $B$ on $\F_4\otimes Y$.
(On the original $\F_4$-space $\F_4^m$, $A$ is a scalar, so this splitting only appears after the extension.)
Let $x$ and $y$ be eigenvectors of $A$ for $\lambda$ and $\mu$.
The $\omega$-eigencomponent of $\beta_\omega(x,y)$ is nonzero only if $\lambda\mu=\omega$, that is, $\lambda=\mu=\omega^2$.
In particular $\beta_\omega$ vanishes on pairs of eigenvectors for $\omega$ and $\omega^2$, since $B$ has no eigenvalue $1$ on $Y$.
The $\omega^2$-eigencomponent is the conjugate of the $\omega$-eigencomponent.
So $\beta_\omega$ is determined by its restriction to the $\omega^2$-eigenspace,
which is an $r$-tuple of $\F_4$-bilinear alternating forms in $m$ variables.
In suitable coordinates, $\beta_\omega(x,y)=(c_1(\bar x,\bar y),\dots,c_r(\bar x,\bar y))$, where $\bar x$ is the coordinatewise Frobenius image of $x$.
Since $m$ is odd, $c_1$ is degenerate. Take $a\neq 0$ with $\bar a$ in its radical.
Then $\beta_\omega(a,\cdot)$ is semilinear over $\F_4$ with image in an $\F_4$-space of dimension at most $r-1$.
So its kernel has dimension at least $2m-2(r-1)$ over $\F_2$.
The $\Fix(B)$-component adds at most $2m-2r$ linear conditions.
So $\dim\ker\beta_a\ge 2$, and $F$ is not APN.
\end{proof}

In dimension ten ($m=5$), Lemma~\ref{lem:odd} excludes types~55 to~59.
In Table~\ref{tab:proofs}, types~57, 58 and~59 are counted under Lemma~\ref{lem:order3}, which also excludes them, and types~55 and~56 under Lemma~\ref{lem:odd}.
Among them, type~56 has a Knuth estimate of about $10^{43}$ nodes.
As a cross-check, an exhaustive search confirms the case $m=3$ in dimension six.

\subsection{Compression}
Lemmas~\ref{lem:intertwiner} and~\ref{lem:dual} follow from one observation.
For $a\neq 0$ and a subspace $D$, let
\[
E(a,D)=\spn\{\beta(a,d) : \beta\in W(A,B),\ d\in D\}.
\]

\begin{proposition}[Compression criterion]\label{prop:compression}
If some $F$ with polar form in $W(A,B)$ is APN, then for every $a\neq 0$ and every subspace $D$,
\[
\dim D-[a\in D]\le\dim E(a,D).
\]
\end{proposition}

\begin{proof}
The map $\beta_a$ sends $D$ into $E(a,D)$, with kernel $\{0,a\}\cap D$.
\end{proof}

Proposition~\ref{prop:compression} is a computation on $W(A,B)$.
By bilinearity, $E(a,D)$ is spanned by the vectors $w_k(a,d_j)$, where $w_k$ runs over a basis of $W(A,B)$ and $d_j$ over a basis of $D$.
A failing pair $(a,D)$ is a certificate that the type is empty.
The certificate is valid only if the basis spans all of $W(A,B)$, not just a subspace.
The artifact therefore stores, for each excluded type, the matrices $A$ and $B$, a basis of $W(A,B)$, and a witness for the first criterion that excludes it.
For this criterion and for the Wong-sequence criterion the witness is the pair $(a,D)$.
For Lemma~\ref{lem:restriction} it is a basis of the subspace, and for Lemma~\ref{lem:order3} the two dimensions it compares.
When $W=0$ or Lemma~\ref{lem:odd} applies, no witness beyond $A$, $B$ and the basis is needed.
An independent checker recomputes the rank of the linear system that defines $W(A,B)$ from $A$ and $B$, checks that the stored basis has the full dimension and satisfies the system, and then checks the witness~\cite{artifact}.
Lemma~\ref{lem:intertwiner}~(i) and~(ii) are the case $a\in\Fix(A)$, $D=\ker g(A)$.
Lemma~\ref{lem:intertwiner}~(iii) is the case $a\in\Fix(A)$, $D=\Img(A+I)$.
Lemma~\ref{lem:dual} is the case $a\in\ker\varphi(A)$, $D=\Fix(A)$.
But the criterion is stronger, since $E(a,D)$ is computed from the actual space $W(A,B)$.
We test all $a\neq 0$ and all sums $D$ of the subspaces $\ker(g(A)^j)$ as above.

For two types the invariant subspaces are not enough, and we choose $D$ adapted to $a$.
Fix $a$ and a generic map $M_0=\beta_a$ for some $\beta\in W(A,B)$.
Let $\mathcal{M}_a=\{\beta_a:\beta\in W(A,B)\}$.
The second Wong sequence~\cite{Wong74,IvanyosKQS15} $N_0=0$, $N_{i+1}=\mathcal{M}_a(M_0^{-1}(N_i))$ is increasing.
Here $M_0^{-1}(N)$ is the full preimage, and $\mathcal{M}_a(D)$ is the span of all $M(d)$ with $M\in\mathcal{M}_a$, $d\in D$; both are computed by linear algebra from bases.
Its limit $N^*$ gives $D=M_0^{-1}(N^*)$ with $\mathcal{M}_a(D)\subseteq N^*$.
We then check Proposition~\ref{prop:compression} for this $D$ directly.
So the result is a certificate even though $M_0$ was chosen at random.
This settles types~130 and~169.

\subsection{Summary of the exclusions}

\begin{table}[t]
\centering
\caption{Types excluded by proof. A type is counted under the first criterion that applies.}\label{tab:proofs}
\begin{tabular}{@{}lr@{}}
\toprule
Criterion & types \\
\midrule
$W(A,B)=0$ & 31 \\
Compression, invariant subspaces (Proposition~\ref{prop:compression}) & 237 \\
Restriction (Lemma~\ref{lem:restriction}) & 4 \\
Order three (Lemma~\ref{lem:order3}) & 4 \\
Compression with Wong sequences & 2 \\
Odd dimension over $\F_4$ (Lemma~\ref{lem:odd}) & 2 \\
\midrule
Total & 280 \\
\bottomrule
\end{tabular}
\end{table}

Table~\ref{tab:proofs} summarizes the exclusions.
Lemma~\ref{lem:intertwiner} and Lemma~\ref{lem:restriction} on $\Fix(A)$ alone already exclude 170 types, using only the matrices $A$ and $B$.
The bound~\eqref{eq:bbf} and Lemma~\ref{lem:dual} exclude no type beyond Table~\ref{tab:proofs}.
No criterion excludes a type in which our searches found an APN function.
This is a useful consistency check of the implementation.
For most excluded types we also ran the generic search of Section~\ref{sec:search}, and it found no APN function.

\section{Exhaustive search}\label{sec:search}

This section describes the generic search, which works for any type.
Section~\ref{sec:special} describes the dedicated methods for types~24 and~47.

\subsection{The search tree}
Fix a type $(A,B)$ and a basis $w_1,\dots,w_d$ of $W(A,B)$.
A polar form in $W(A,B)$ is $\beta=\sum_k t_kw_k$ with $t\in\F_2^d$.
The 450 unknowns are the coordinates $\beta(e_i,e_j)_r$, ordered first by the pair $(i,j)$ in the order
$(0,1),(0,2),(1,2),(0,3),\dots$ and then by the output coordinate $r=0,\dots,9$.
We put the basis in reduced echelon form with respect to this order, with the pivot of $w_k$ at its first nonzero unknown.
We say that the pivot of $w_k$ is at the pair $(i,j)$ of that unknown.
Then the value $\beta(e_i,e_j)$ depends only on the coefficients $t_k$ whose pivot is at or before the pair $(i,j)$.

The search follows the variable approach of earlier work~\cite{DBLP:journals/dcc/YuWL14,DBLP:journals/tit/BeierleL22}.
Level $s$ adds the input vector $e_s$.
Within a level, the pairs $(0,s),(1,s),\dots,(s-1,s)$ are filled in order.
At each pair, the coefficients with a pivot there are chosen in all possible ways.
Let $V_s=\spn\{e_0,\dots,e_{s-1}\}$.
By Lemma~\ref{lem:apn-kernel}, $F$ is APN on $V_{s+1}$ if and only if it is APN on $V_s$ and
\begin{equation}\label{eq:variable}
\beta(a,e_s)\notin\beta_a(V_s)\qquad\text{for every }a\in V_s,\ a\neq 0.
\end{equation}
Indeed, let $x,y\in V_{s+1}$ be linearly independent.
If both lie in $V_s$, the condition on $V_s$ applies.
If exactly one, say $x$, lies in $V_s$, write $y=v+e_s$ with $v\in V_s$; then $\beta(x,y)=\beta_x(v)+\beta(x,e_s)$, which is nonzero by~\eqref{eq:variable} with $a=x$.
If neither lies in $V_s$, then $x+y\in V_s$ is nonzero and $\beta(x,y)=\beta(x,x+y)$, which is the previous case.
Condition~\eqref{eq:variable} is tested for $a$ as soon as $\beta(e_i,e_s)$ is known for the highest index $i$ in $a$.
We keep a reduced echelon basis of each space $\beta_a(V_s)$, so each test takes at most $n$ steps.
A leaf of the tree is a quadratic APN function of $\F_2^{10}$ with the self-equivalence $(A,B)$.

\subsection{Using the centralizers}
The centralizers $\mathcal{C}(A)$ and $\mathcal{C}(B)$ act on $W(A,B)$ by $\beta\mapsto S\beta(T\cdot,T\cdot)$ with $T\in \mathcal{C}(A)$, $S\in \mathcal{C}(B)$.
This action preserves the EA-class.
We use it in three ways, and each one keeps at least one member of every orbit.
\begin{itemize}
\item \emph{First value.}
The value $\beta(e_0,e_1)$ is restricted to representatives of the orbits of $\mathcal{C}(B)$ on nonzero vectors.
\item \emph{Output side.}
On the first pairs, the search keeps only values that are lexicographically least under a stabilizer chain in $\mathcal{C}(B)$.
This comparison uses a different order from the echelon form above.
Each value $\beta(e_i,e_j)\in\F_2^{10}$ is read as an integer, and sequences of values are compared pair by pair with the usual order of integers.
If a value ties with its image, the value is kept and the comparison moves to the next pair.
\item \emph{Input side.}
At the end of some levels, we apply sampled maps $T\in \mathcal{C}(A)$ that preserve $V_s$, followed by the output reduction.
If the image of the current prefix is lexicographically smaller, the branch is pruned.
The order is the same lexicographic order on the sequence of values, and $T$ preserves $V_s$.
So if the image of a prefix is smaller, the image of every completion is smaller too.
Hence the least member of each orbit is never pruned, and the reduction is sound even with sampled maps.
\end{itemize}
For several hard types it also helps to reorder the input basis:
first $\Img(A+I)$, then $\Fix(A)$, then a complement.
The random choices are reseeded from a hash of the current prefix.
So the pruning is deterministic, and the tree can be split into independent shards that are run in parallel.

\subsection{Cost}
We estimate the size of each tree with Knuth's method~\cite{Knuth75}, before and after the reductions.
Each estimate averages 2,000 or 3,000 random root-to-leaf probes with a fixed seed.
These estimates are heuristic; we use them only to choose a method, never as part of a proof.
For example, type~18 goes from $1.8\cdot 10^{44}$ to $1.9\cdot 10^{8}$ nodes.
The generic search completed on 258 types.
The classification and confirmation searches took about 36,400 core-hours: 13,300 on type~47, 12,600 on type~73, 7,500 on type~24, and about 3,000 on all other searched types.
The last figure includes the confirmation searches on the 242 types that are also excluded by proof.
It does not count the type-64 searches or the Knuth estimates.
The runs used three Linux servers (Ubuntu 22.04).
Each has two AMD EPYC 7763 processors (64 cores each, 2.45\,GHz, two hardware threads per core, so 256 threads) and 1\,TB of RAM.
The servers were shared with other users, and our jobs could use at most 200 threads per server.
Every job is single-threaded and ran at the lowest priority.
A core-hour here is one hour of wall-clock time of one job on one hardware thread.
It is not a measurement on a dedicated core.
The search engine is written in C and compiled with gcc~11.4 and \texttt{-O3}.
The equivalence tests are in Python.
Each type's cost is listed in the artifact~\cite{artifact}.

\subsection{Validation}
We validated the search on cases with a known answer.
\begin{itemize}
\item The space of all functions $cx^3$ gives exactly 1023 APN functions, and one after normalization.
The space of all $cx^5$ gives none.
\item Type~231 (order 31) gives Beierle--Leander \#1~\cite{DBLP:journals/tit/BeierleL22}.
\item In dimension eight, type~68 of the dimension-eight list (order 7) contains 9 classes.
The search without reductions ($7.9\cdot10^6$ nodes) and with both reductions ($2.9\cdot 10^4$ nodes) give the same 9 classes.
The same holds when the run is split into shards.
\item In dimension ten, type~176 gives the same 11 classes with and without the reductions.
\item Splitting a run into shards and running the same program again tests the splitting, not the reductions.
The tests with and without reductions, and the cases with a known answer, check the implementation of the reductions.
They complement the soundness arguments above; they do not replace them.
\end{itemize}
The APN property of every reported function, including the representatives of the four new classes, is checked independently from its full difference table.

\subsection{Classifying the leaves}\label{sec:ea}
Every leaf is matched to a known class by an explicit EA-equivalence.
We first compute the ortho-derivative invariants of the leaf~\cite{DBLP:journals/tit/CanteautCP22}.
They select a known representative $F$ with the same invariants, or show that the leaf is new.
The certificate search is exhaustive: if it finds no equivalence, the leaf is not EA-equivalent to $F$, and we keep it as a possible new class.
This never happened: every leaf whose invariants matched a known class was certified equivalent to it.
Then we search for linear $L_1,L_2$ and affine $L_3$ with $G=L_1\circ F\circ L_2+L_3$.
The search builds $L_2$ one basis vector at a time and derives $L_1$ by linear algebra.
Two observations reduce the cost of a test.
\begin{itemize}
\item \emph{Kernel partners.}
For a quadratic APN function, every nonzero linear form $v$ defines a component whose polar form has a radical $K^F_v$.
If $G=L_1\circ F\circ L_2+L_3$, then $K^G_v=L_2^{-1}\bigl(K^F_{L_1^{\top}v}\bigr)$.
So $L_2$ maps the points that share a radical with $y$ onto the points that share a radical with $L_2y$.
The search chooses the next basis vector among such partners whenever possible.
Its image is then restricted to the partners of the image.
\item \emph{Transitivity.}
For the Gold functions, the maps $x\mapsto ax$ act transitively on the nonzero inputs.
So the image of the first basis vector can be fixed.
\end{itemize}
Each certificate $(L_1,L_2,L_3)$ is checked independently by recomputing $L_1\circ F\circ L_2+L_3$.
With these two observations, a test takes between 0.2 and 1.4 seconds in Python on one core.
Without them, a test against the Gold function $x^3$ took more than ten minutes.

\subsection{Results of the generic search}

\begin{table}[t]
\centering
\caption{Types in which the generic search finds APN functions.
Leaves are APN functions in the search space after normalization.}\label{tab:generic}
\begin{tabular}{@{}rrrll@{}}
\toprule
Type & $p$ & leaves & classes & names \\
\midrule
129 & 11 & 250       & 4  & Gold $x^3$, $x^9$, Beierle--Leander \#5, F13 \\
170 & 31 & 338\,272  & 1  & F3 \\
171 & 31 & 338\,272  & 1  & F3 \\
175 & 31 & 492\,032  & 1  & Gold $x^3$ \\
176 & 31 & 3\,189\,280 & 11 & Beierle--Leander \#3, six F12, four F13 \\
180 & 31 & 492\,032  & 1  & Gold $x^9$ \\
231 & 31 & 961       & 1  & Beierle--Leander \#1 \\
73  & 5  & 4\,536    & 24 & see Section~\ref{sec:t73} \\
\bottomrule
\end{tabular}
\end{table}

Table~\ref{tab:generic} lists the types in which the generic search finds APN functions.
Every leaf was tested.
Each one is EA-equivalent to a known class or to one of the four new classes of type~73.
The other types searched by the generic method contain no APN function.
Eight of them are not excluded by any lemma of Section~\ref{sec:lemmas}:
types~116, 132, 198, 207, 208, 209, 235 and~255.
For these, as for type~47 in Section~\ref{sec:special}, the search is the only proof.

\section{Three hard types}\label{sec:special}

Three types not excluded by proof are too large for a direct search with the basic settings:
types~73, 24 and~47.
For type~73 the generic search with all reductions is enough, on a large scale.
Types~24 and~47 need a normal form and a dedicated solver.

\subsection{Type 73: coefficients in $\F_4$}\label{sec:t73}
Type~73 is the set of quadratic functions of $\F_{2^{10}}$ with coefficients in $\F_4$, up to EA-equivalence
(Section~\ref{sec:types}).
This is the case $(n,m)=(10,2)$ of the search with subfield coefficients of Ichanska et al.,
who report it as infeasible~\cite{DBLP:journals/ccds/IchanskaBKY26}.
Beierle and Leander ran a randomized, non-exhaustive search over most types in dimension ten~\cite{DBLP:journals/tit/BeierleL22}.
Two of their new functions, Beierle--Leander \#1 and \#5, lie in type~73.

We split the tree into 192 shards.
The run explored about $1.9\cdot 10^{12}$ nodes and took 12,600 core-hours.
A second run, split into 3,072 shards, printed every leaf.
Both runs agree piece by piece (16 shards of the second run for each of the 192 pieces of the first).
They have the same number of nodes and the same 4,536 leaves.
Every leaf was matched to a class by an exact certificate (Section~\ref{sec:ea}).

\begin{theorem}[computer-assisted]\label{thm:t73}
Up to EA-equivalence, there are exactly 24 quadratic APN functions of $\F_{2^{10}}$ with all coefficients in $\F_4$.
Twenty of them are published classes. Four are new.
\end{theorem}

Type~73 contains 20 published classes.
Nine of them are the Gold functions $x^3$ and $x^9$,
Beierle--Leander \#1 and \#5~\cite{DBLP:journals/tit/BeierleL22},
and the triplicate functions \#1, \#3, \#4, \#5 and \#6 of Budaghyan, Ivkovic and Kaleyski~\cite{DBLP:journals/ccds/BudaghyanIK23}.
The other eleven come from known infinite families~\cite{DBLP:journals/ffa/BudaghyanCV20}.
We call these families F3, F4, F12 and F13, as in the artifact~\cite{artifact}.
Type~73 contains two, two, two and five classes of them.
The five published classes missing from type~73 are Beierle--Leander \#3 and four F12 classes.
They all lie in type~176.

\begin{table}[t]
\centering
\caption{The four new quadratic APN classes of $\F_2^{10}$.
$M$ is the largest dimension of a space of non-bent components.
The spectrum is the differential spectrum of the ortho-derivative, written as value$^{\text{multiplicity}}$.}\label{tab:new}
\begin{tabular}{@{}clc@{}}
\toprule
Class & ortho-derivative differential spectrum & $M$ \\
\midrule
N1 & $0^{635846}\,2^{317181}\,4^{79085}\,6^{13540}\,8^{1725}\,10^{145}\,12^{30}$ & 3 \\
N2 & $0^{633636}\,2^{320446}\,4^{79035}\,6^{12670}\,8^{1580}\,10^{180}\,12^{5}$ & 4 \\
N3 & $0^{635924}\,2^{316881}\,4^{79575}\,6^{13245}\,8^{1660}\,10^{237}\,12^{25}\,14^{5}$ & 4 \\
N4 & $0^{634720}\,2^{318706}\,4^{79315}\,6^{13035}\,8^{1575}\,10^{176}\,12^{20}\,14^{5}$ & 4 \\
\bottomrule
\end{tabular}
\end{table}

Table~\ref{tab:new} lists the four new classes.
Their ortho-derivative spectra differ from those of all published classes and from each other.
So they are four new EA-classes, hence four new CCZ-classes.
We use the conventions of~\cite{DBLP:journals/tit/CanteautCP22}, and we reproduced the spectra published in~\cite{HellesethIK26}.
We computed $M$ by listing all linear subspaces of $\mathcal{N}$ of dimension up to five; the artifact recomputes it.
All four classes have $M<5$, so none of them is CCZ-equivalent to a permutation.
Each new class has a univariate representative with 24 to 26 terms and all coefficients in $\F_4$.
These representatives are listed in Appendix~\ref{app:new} and in the artifact~\cite{artifact}.

\subsection{Type 24: coefficients in $\F_{32}$}\label{sec:t24}
Type~24 is the set of quadratic functions of $\F_{2^{10}}$ with coefficients in $\F_{32}$.
It has a special role for the big APN problem.
In dimension six, the quadratic APN functions with coefficients in $\F_8$ form exactly two EA-classes: the Gold class and the class of Kim's function.
Kim's function is $x^3+x^{10}+gx^{24}$ on $\F_{64}$, with $g$ a primitive element.
Its coefficients are not in $\F_8$, but its EA-class contains $ax^5+ax^{33}+x^{40}$ with $a=g^{54}\in\F_8$, where $g$ is a root of $x^6+x^4+x^3+x+1$; we checked an explicit certificate.
Kim's function is CCZ-equivalent to Dillon's APN permutation~\cite{BrowningDMW10}.
The next lemma shows that the bound $M\ge n/2$ never rules out this case.

\begin{lemma}\label{lem:subfield}
Let $n=2m$ with $m$ odd, and let $F$ be a quadratic function of $\F_{2^n}$ with all coefficients in $\F_{2^m}$.
Then the components $\Tr(cF)$ with $c\in\F_{2^m}^*$ are not bent.
In particular $M\ge n/2$.
\end{lemma}

\begin{proof}
Let $A=B$ be the Frobenius $x\mapsto x^{2^m}$, so that~\eqref{eq:equivariant} holds.
Put $\epsilon=A+I$.
Expanding $\beta(Ax,Ay)=B\beta(x,y)$ gives
$\beta(\epsilon x,y)+\beta(x,\epsilon y)+\beta(\epsilon x,\epsilon y)=\epsilon\beta(x,y)$.
With $x=\epsilon u$ and $\epsilon^2=0$ this becomes $\beta(\epsilon u,\epsilon y)=\epsilon\beta(\epsilon u,y)$.
Let $c$ be a linear form with $c\circ\epsilon=0$; these are the forms $\Tr(c\,\cdot)$ with $c\in\F_{2^m}$.
Put $b=c\circ\beta$ and $\gamma(u,v)=b(\epsilon u,v)$.
Then:
(a) $b(\epsilon u,\epsilon v)=0$; since $\ker\epsilon=\Fix(A)$, this shows that $\gamma$ is defined on the quotient $\F_{2^n}/\Fix(A)$ in both arguments, and this quotient has dimension $m$;
(b) $\gamma(u,v)+\gamma(v,u)=b(\epsilon u,\epsilon v)=0$ by the same identity, so $\gamma$ is symmetric;
(c) $\gamma(u,u)=c(\beta(\epsilon u,u))=0$, since $\beta(\epsilon u,u)=\beta(Au,u)$ lies in $\Fix(B)$, and $\Fix(B)=\Img\epsilon\subseteq\ker c$.
So $\gamma$ is alternating in odd dimension $m$, hence degenerate.
Take $u\notin\Fix(A)$ in its radical. Then $\epsilon u\neq 0$ is in the radical of $b$.
\end{proof}

So the simplest necessary condition for a permutation holds for every function of type~24.
This makes type~24 the place to look for an analogue of Kim's function.
Its search tree is large: about $9\cdot 10^{26}$ nodes for the generic search, even with a fixed Gold prefix.
We use a normal form instead.

Write $x=(s,t)$ with $s,t\in\F_2^5$, and choose bases such that $A(s,t)=(s,s+t)$ and $B=A$ on the output.
Then $\Fix(A)=\{(0,t)\}$.

\begin{lemma}[Normal form]\label{lem:form24}
Every quadratic APN function with a self-equivalence of type~24 is EA-equivalent to a function
\begin{equation}\label{eq:form24}
F(s,t)=\bigl(G(s)+\nu(s,s)+\mu(s,t),\; R(s)+\nu(t,s)+G(t)\bigr),
\end{equation}
where $G$ is the Gold function $x^3$ or $x^5$ of $\F_{32}$, $\mu$ is the polar form of $G$,
$\nu$ is a bilinear map, and $R$ is a quadratic map with polar form $\rho$.
Conversely, every function~\eqref{eq:form24} has the self-equivalence $(A,A)$.
\end{lemma}

\begin{proof}
Write $\beta=(\beta_1,\beta_2)$ for the two output halves, and
$\beta_i((a,b),(u,c))=X_i(a,u)+Y_i(a,c)+Y_i(u,b)+Z_i(b,c)$ with bilinear $X_i,Y_i,Z_i$.
The equation $\beta(Ax,Ay)=A\beta(x,y)$ has two halves.
The first half reads $\beta_1(Ax,Ay)=\beta_1(x,y)$.
Expanding, it gives $Z_1=0$ and $Y_1(a,u)+Y_1(u,a)=0$.
The second half reads $\beta_2(Ax,Ay)+\beta_2(x,y)=\beta_1(x,y)$.
Expanding, it gives $Y_1=Z_2$ and $X_1(a,u)=Y_2(a,u)+Y_2(u,a)+Z_2(a,u)$.
The maps $X_2$, $Y_2$ and $Z_2$ are free.

The restriction $t\mapsto F(0,t)$ has polar form $(0,Z_2)$.
By Lemma~\ref{lem:apn-kernel}, $Z_2$ is the polar form of a quadratic APN function of $\F_2^5$.
By the classification of the APN functions of $\F_2^5$~\cite{BrinkmannL08}, a quadratic one is EA-equivalent to $x^3$ or $x^5$.
The maps $(s,t)\mapsto(Ls,Lt)$ with $L\in\GL(5,2)$ commute with $A$.
Applying such maps on the input and the output, we may assume $Z_2=\mu$.
Put $\nu(c,u)=Y_2(u,c)$ and $\rho=X_2$.
Then $\beta_1((a,b),(u,c))=\mu(a,u)+\nu(a,u)+\nu(u,a)+\mu(a,c)+\mu(b,u)$
and $\beta_2((a,b),(u,c))=\rho(a,u)+\nu(b,u)+\nu(c,a)+\mu(b,c)$.
These are the polar forms of~\eqref{eq:form24}.
Two quadratic functions with the same polar form differ by an affine map, which proves the first claim.
The converse is the same computation read backwards.
\end{proof}

Lemma~\ref{lem:form24} leaves the bilinear map $\nu$ (125 bits) and the alternating form $\rho$ (50 bits).
The next lemma turns APN into one condition per plane on $\rho$.
A \emph{plane} is a 2-dimensional subspace of $\F_2^5$.

\begin{lemma}[Planes]\label{lem:planes}
Let $F$ be as in~\eqref{eq:form24}.
For linearly independent $u,v\in\F_2^5$, let $\Sigma(u,v)$ be the set of values $\nu(b,v)+\nu(c,u)+\mu(b,c)$,
over all $b,c\in\F_2^5$ with $\mu(u,v)+\nu(u,v)+\nu(v,u)+\mu(u,c)+\mu(b,v)=0$.
For a plane $P$, let $\mathrm{Forb}(P)$ be the union of $\Sigma(u,v)$ over the three bases $\{u,v\}$ of $P$.
Then $F$ is APN if and only if
\begin{enumerate}
\item[(i)] $\nu(b,b)\notin\Img\mu(b,\cdot)$ for every $b\neq 0$, and
\item[(ii)] $\rho(u,v)\notin\mathrm{Forb}(P)$ for every plane $P=\langle u,v\rangle$.
\end{enumerate}
\end{lemma}

\begin{proof}
Since $\rho$ is alternating, $\rho(u,v)=\rho(u,u+v)=\rho(u+v,v)$, so $\rho(u,v)$ depends only on $P$.
By Lemma~\ref{lem:apn-kernel}, $F$ is APN if and only if $\beta(x,y)\neq 0$ for all linearly independent $x=(a,b)$ and $y=(u,c)$.
We use the formulas for $\beta_1$ and $\beta_2$ from the proof of Lemma~\ref{lem:form24}.

\emph{Case $a=u=0$.} Then $\beta(x,y)=(0,\mu(b,c))$, which is nonzero since $G$ is APN.

\emph{Case $a=0$, $u\neq 0$.} Then $b\neq 0$, and $\beta_1=\mu(b,u)$ vanishes only for $u=b$.
For $u=b$ we get $\beta_2=\nu(b,b)+\mu(b,c)$.
So $\beta(x,y)\neq 0$ for all such $y$ if and only if (i) holds for $b$.
The case $a\neq 0$, $u=0$ is the same, since $\beta$ is symmetric.

\emph{Case $a=u\neq 0$.} Then $\beta(x,y)=\beta(x,x+y)$, and $x+y=(0,b+c)$.
This is the previous case.

\emph{Case $a,u$ independent.} They span a plane $P$, and $\beta_2=\rho(a,u)+\nu(b,u)+\nu(c,a)+\mu(b,c)$.
So $\beta(x,y)=0$ for some $b,c$ if and only if $\rho(a,u)\in\Sigma(a,u)$.
Since $\beta$ is symmetric, $\Sigma(a,u)=\Sigma(u,a)$.
The ordered pairs $(a,u)$ of independent vectors of $P$ give the three bases of $P$.
So this case is exactly (ii).
\end{proof}

Lemma~\ref{lem:planes} turns APN into one ``avoid a set'' condition per plane on $\rho$.
There are 155 planes in $\F_2^5$.
The search uses two groups of transformations that preserve the normal form~\eqref{eq:form24} and the EA-class.

\emph{Translations of $\nu$.}
For $Y,Y'\in M_5(\F_2)$, the maps $(s,t)\mapsto(s,t+Ys)$ on the input and $(u,v)\mapsto(u,v+Y'u)$ on the output commute with $A$.
They keep $G$ and $\mu$, change $\rho$, and replace $\nu$ by $\nu+\theta$ with $\theta(x,y)=\mu(x,Yy)+Y'\mu(x,y)$.
These $\theta$ form a space $\mathcal{T}$ of dimension 49.
So only the coset $\nu+\mathcal{T}$ matters.
We take the unique element of the coset that vanishes on the pivot coordinates of a reduced echelon basis of $\mathcal{T}$.
This leaves 45 free bits for $\nu$ after condition~(i) of Lemma~\ref{lem:planes}.
Under a translation, $(b,c)\mapsto(b+Yu,c+Yv)$ is a bijection of the solutions in the definition of $\Sigma(u,v)$, and $\Sigma(u,v)$ is shifted by a vector that depends only on the plane $\langle u,v\rangle$.

\emph{Automorphisms of $G$.}
The maps $L(x)=\lambda x^{2^k}$ ($\lambda\in\F_{32}^*$, $0\le k<5$) satisfy $G\circ L=N\circ G$ for a linear $N$.
Such a map sends $F$ to $F'=\mathrm{diag}(N^{-1},N^{-1})\circ F\circ\mathrm{diag}(L,L)$.
Both coordinate changes commute with $A$, and $F'$ keeps $\mu$, since $N^{-1}\mu(Lx,Ly)=\mu(x,y)$.
It replaces $\nu$ by $\nu'(x,y)=N^{-1}\nu(Lx,Ly)$.
Hence $\mathrm{Forb}_{F'}(P)=N^{-1}\mathrm{Forb}_F(LP)$, with the three parts permuted.

\begin{lemma}\label{lem:regular}
The 155 maps $x\mapsto\lambda x^{2^k}$ act regularly on the 155 planes of $\F_{32}$.
\end{lemma}

\begin{proof}
It suffices to show that only the identity fixes a plane $P$.
Let $g\neq\mathrm{id}$ fix $P$. The group has order $155$, which is prime to $3!$.
So $g$ fixes each of the three nonzero points of $P$.
If $k=0$, then $\lambda a=a$ for $a\neq 0$, so $\lambda=1$.
If $k\neq 0$, the nonzero fixed points satisfy $x^{2^k-1}=\lambda^{-1}$.
Since $\gcd(2^k-1,31)=1$, this has exactly one solution, so $g$ has at most one nonzero fixed point.
\end{proof}

\emph{The key of a plane.}
For each plane $P$ we compute a key: the size of $\mathrm{Forb}(P)$, the sorted sizes of its three parts, and, to break ties, the affine dimension, the sizes of the pairwise and triple intersections of the parts, and a hash of their multiplicity histograms.
By the two previous paragraphs, the key of $P$ is unchanged by translations of $\nu$, and $\mathrm{key}_{F'}(P)=\mathrm{key}_F(LP)$ after an automorphism.
Moreover $\mathrm{Forb}(P)$ depends only on the columns $\nu(\cdot,w)$ with $w\in P$.
So the keys of the planes inside $\spn\{e_0,\dots,e_j\}$ are fixed once the first $j+1$ columns of $\nu$ are chosen.

\emph{The solver.}
It runs for $G=x^3$ and $G=x^5$ separately.
\begin{enumerate}
\item Choose the columns $\nu(\cdot,e_0)$, $\nu(\cdot,e_1)$, $\nu(\cdot,e_2)$ in reduced form, each satisfying condition~(i).
After columns~2 and~3, reject the prefix if some plane inside the current span has a strictly smaller key than $\langle e_0,e_1\rangle$.
\item List all values of $\rho$ on $\spn\{e_0,e_1,e_2\}$ that satisfy condition~(ii) on its planes.
\item Choose column $\nu(\cdot,e_3)$, apply the key test, and extend $\rho$ to $\spn\{e_0,\dots,e_3\}$.
\item For each option of column $\nu(\cdot,e_4)$, extend $\rho$ to $\F_2^5$ and check condition~(ii) on the remaining planes.
\item Every complete $(\nu,\rho)$ is a leaf. Its lookup table is rebuilt from~\eqref{eq:form24} and checked to be APN.
\end{enumerate}

\emph{Coverage.}
Let $F$ be APN of type~24.
By Lemma~\ref{lem:form24}, it is EA-equivalent to a function~\eqref{eq:form24}.
Choose a plane $P_0$ of smallest key.
By Lemma~\ref{lem:regular}, there is exactly one automorphism $L$ of $G$ with $L\langle e_0,e_1\rangle=P_0$.
For the resulting $F'$ we have $\mathrm{key}_{F'}(\langle e_0,e_1\rangle)=\mathrm{key}_F(P_0)\le\mathrm{key}_F(LP)=\mathrm{key}_{F'}(P)$ for every plane $P$.
Then reduce $\nu$ modulo $\mathcal{T}$; this does not change any key.
The result has $\langle e_0,e_1\rangle$ of smallest key among all planes.
So it passes every key test, since the tests reject only strictly smaller keys.
It satisfies conditions~(i) and~(ii), so it is a leaf.
Hence every class of type~24 has at least one leaf.

The run is split into $4{,}096$ jobs: for each $G$, job $r\in\{0,\dots,2047\}$ keeps the options of column~1 whose index is congruent to $r$ modulo 2048.
The artifact lists the expected job identifiers and the completion record of each job, and a script checks that all of them finished with exit code zero.
The full run took about 7,500 core-hours.
It printed every leaf: there are three.
In our runs, contradictions appeared only at the last step, when $\rho$ is extended to all of $\F_2^5$.

We validated the solver in two ways.
First, an earlier implementation of the same method finds exactly the Gold and Kim classes in dimension six.
Second, we planted known functions of type~24 in all 155 positions given by the automorphisms of $G$.
The production solver recovers $x^3$, $x^9$ and $x^3+\Tr(x^9)$ from every position allowed by the key rule.

\begin{theorem}[computer-assisted]\label{thm:t24}
Up to EA-equivalence, there are exactly three quadratic APN functions of $\F_{2^{10}}$ with all coefficients in $\F_{32}$:
$x^3$, $x^9$, and $x^3+\Tr(x^9)$.
\end{theorem}

All three have $M=5$. Since $\F_2\subset\F_{32}$, Theorem~\ref{thm:t24} also covers the quadratic APN functions with coefficients in $\F_2$.
All three classes of Theorem~\ref{thm:t24} have representatives with coefficients in $\F_2$.
So there are exactly three such classes.
This agrees with Yu et al., who also found three classes with a different method~\cite{YuLIK25}.
Only the two Gold functions pass the counting criterion, and they fail the projection test~\cite{Bellini26ccz}.
So, unlike dimension six, type~24 contains no function CCZ-equivalent to a permutation.

\subsection{Type 47}\label{sec:t47}
In type~47, $A=B$ has order three, with $U=\ker(A^2+A+I)\cong\F_4^3$ and $Z=\Fix(A)$ of dimension four.
Equivariance leaves four blocks of $\beta$.
The block on $U\times U$ is a pair $(\kappa,h)$, with $\kappa$ into $U$ and $h$ into $Z$.
The block on $U\times Z$ is an intertwiner $\tau(u,z)=T_zu$ into $U$, which is $\F_4$-linear in $u$.
The block on $Z\times Z$ is the polar form $Q$ of a quadratic function from $Z$ to $Z$.

\begin{lemma}\label{lem:rankh}
If a function of type~47 is APN, then for every $u\in U$, $u\neq 0$:
$\rank h(u,\cdot)\ge 3$, and $\{x\in U : h(u,x)=0,\ \kappa(u,x)=0\}=\{0,u\}$.
\end{lemma}

\begin{proof}
The $Z$-part of $\beta((u,0),\cdot)$ is $h(u,\cdot)$, and its $U$-part has rank at most $\dim U=6$.
APN needs rank $9$, so $\rank h(u,\cdot)\ge 3$.
The second condition is Lemma~\ref{lem:apn-kernel} restricted to $U$.
\end{proof}

We now describe the search.
Identify $U$ with $\F_4^3$, with $A$ acting as $\omega$.
The block $h$ has four $\F_2$-valued coordinates, and each is of the form $h_i(x,y)=\Tr_{4/2}(x^*M_iy)$ with $M_i$ a $3\times 3$ Hermitian matrix over $\F_4$.
Up to linear maps on the output $Z$, only the span $S$ of $M_1,\dots,M_4$ matters.
The block $\kappa$ is $\F_4$-bilinear in $(\bar x,\bar y)$ and alternating, so it factors through the cross product: $\kappa(x,y)=K^{\top}(\bar x\times\bar y)$ for a $3\times 3$ matrix $K$ over $\F_4$.
An $\F_4$-linear output map $R$ sends $K$ to $KR^{\top}$, so only the column space $V$ of $K$ matters.
Since $(\bar P\bar x)\times(\bar P\bar y)=\det(\bar P)\,\bar P^{-\top}(\bar x\times\bar y)$, an input map $P$ sends $K$ to $\det(\bar P)\,\bar P^{-1}K$.
An input map $P\in\GL(3,4)$ acts by $S\mapsto P^*SP$ and $V\mapsto\bar P^{-1}V$.
For $x\in U$ we have $\beta(x,\omega x)=(0,h(x,\omega x))$, and $h_i(x,\omega x)$ vanishes for all $i$ exactly when $x^*M_ix=0$ for all $i$.
So APN requires $S$ to be \emph{anisotropic}: $x^*Mx=0$ for all $M\in S$ only if $x=0$.

\begin{enumerate}
\item \emph{Pairs $(S,V)$.}
We list every subspace $S\subseteq\Herm_3(\F_4)$ of dimension at most four, keep the anisotropic ones, and sort them into $\GL(3,4)$-orbits.
Each orbit is checked to be complete: we compute the stabilizer of $S$ by brute force over all $181{,}440$ elements of $\GL(3,4)$, and check $|\mathrm{Stab}(S)|\cdot|\mathrm{orbit}|=|\GL(3,4)|$.
This gives 34 orbits of $S$.
For each, we split the $\F_4$-subspaces $V$ of $\F_4^3$ into orbits of $\mathrm{Stab}(S)$.
This gives 806 classes of pairs $(S,V)$ with $S$ anisotropic.
\item \emph{Lemma~\ref{lem:rankh}.} It removes 703 classes; 103 remain.
\item \emph{Normal form of $\tau$.}
Fix the point $f_0=(1,0,0)\in U$.
The map $z\mapsto\tau(f_0,z)$ is injective, since $\beta(f_0,z)\neq 0$ for $z\neq 0$.
Its image $X$ is a 4-dimensional $\F_2$-subspace of $U$, and APN at $f_0$ requires $X\cap\kappa(f_0,\ker h(f_0,\cdot))=0$.
Input maps on $Z$ act freely on the bases of $X$, so we may choose the basis $z_1,\dots,z_4$ of $Z$ such that $\tau(f_0,z_i)$ is the reduced echelon basis of $X$.
The $\F_4$-linear output maps that fix $\kappa$ act on the possible $X$; we keep one representative per orbit, the one with the smallest bitset.
A \emph{slice} is a class of step~2 together with one representative $X$. There are $8{,}056$ slices.
\item \emph{Completion.}
In each slice, the solver enumerates all remaining values of $\tau$ and all $Q$, testing the APN condition as it goes, as in Section~\ref{sec:search}.
\end{enumerate}

\emph{Coverage.}
Every APN function of type~47 is EA-equivalent, by maps that commute with $A$, to one whose pair $(S,V)$ is one of the 806 representatives.
By Lemma~\ref{lem:rankh} it is one of the 103.
Changing the basis of $Z$ and applying an output map that fixes $\kappa$, we reach a slice.
The solver is exhaustive within a slice.

The solver was validated on the analogue in dimension six, with $U\cong\F_4^2$ and $\dim Z=2$.
There we fixed the blocks $(\kappa,h)$ of $x^3$ and compared with a brute force over all $2^{18}$ completions $(\tau,Q)$.
The brute force finds 144 APN completions, and 6 after the same normalization of $\tau$; the solver finds the same 6, all with the ortho-derivative spectrum of $x^3$.
The artifact lists the 806 classes, the 103 survivors and the 8,056 slices, and a script recomputes the split and checks that every slice finished with no APN function.
All 8,056 slices finished. They took about 13,300 core-hours.
No choice of $(\tau,Q)$ reaches the last level.

\begin{theorem}[computer-assisted]\label{thm:t47}
No quadratic APN function of $\F_2^{10}$ has a self-equivalence of type~47.
\end{theorem}

\section{The open type: canonical triplicate functions}\label{sec:type64}

In type~64, $A$ is the multiplication by $\omega$ on $\F_2^{10}=\F_4^5$ and $B=I$.
So the functions satisfy $F(\omega x)=F(x)$, up to affine terms.
Up to EA-equivalence, these are the quadratic canonical triplicate functions of
Budaghyan, Ivkovic and Kaleyski~\cite{DBLP:journals/ccds/BudaghyanIK23}.
Baudrin, Canteaut and Perrin study them as a family of cyclotomic mappings~\cite{BaudrinCP26}.
No lemma of Section~\ref{sec:lemmas} applies, and the type is too large for our searches (Section~\ref{sec:t64hard}).
It is the only type that we cannot settle.
In this section we give a reformulation, the evidence we have, and the reasons why the type is hard.

\subsection{A Hermitian reformulation}
Write $x^*$ for the conjugate transpose of $x\in\F_4^5$, where conjugation is $z\mapsto z^2$.
Let $\Herm_5(\F_4)$ be the space of $5\times 5$ Hermitian matrices over $\F_4$.
It is a vector space of dimension 25 over $\F_2$.
For $H\in\Herm_5(\F_4)$, the map $x\mapsto x^*Hx$ takes values in $\F_2$, and it is invariant under $x\mapsto\omega x$.
Its polar form is $(q,x)\mapsto\Tr_{4/2}\bigl((Hq)^*x\bigr)$, where $\Tr_{4/2}(z)=z+z^2$.
This polar form is zero only for $H=0$.
So $H\mapsto x^*Hx$ is injective modulo affine terms.
The quadratic Boolean functions invariant under $x\mapsto\omega x$, modulo affine terms, also form a space of dimension~25:
type~64 has $\dim W(A,B)=250$, that is, 25 for each of the ten output coordinates.
So these Boolean functions are exactly the Hermitian forms $x^*Hx$, up to affine terms.
So a function of type~64 is given by ten Hermitian matrices $H_1,\dots,H_{10}$:
\[
F(x)=(x^*H_1x,\dots,x^*H_{10}x).
\]
Up to EA-equivalence, only their $\F_2$-span $V$ matters.
If $F$ is APN, then $\dim V=10$.
Indeed, the coordinates of $F$ are linearly independent modulo affine functions exactly when the values of $\beta$ span $\F_2^{10}$.
They span a space of dimension at least $9$, since $\beta_a$ has rank $9$ for $a\neq 0$, and Lemma~\ref{lem:restriction} with $D=\F_2^{10}$ excludes dimension~9.
On $\Herm_5(\F_4)$ we use the nondegenerate $\F_2$-bilinear form $\langle H,K\rangle=\operatorname{tr}(HK)$,
where $\operatorname{tr}$ is the matrix trace, and we write $C=V^\perp$.
Note that $x^*Hx=\langle H,xx^*\rangle$.

\begin{proposition}\label{prop:herm}
Let $F$ be of type~64 with $\dim V=10$. The following are equivalent.
\begin{enumerate}
\item $F$ is APN.
\item Every point of $\PG(4,4)$ is annihilated by exactly one nonzero matrix of $V$.
\item $F$ is 3-to-1 on $\F_4^5\setminus\{0\}$, with nonzero values.
\item The space $C$ contains no nonzero matrix of rank at most two.
\end{enumerate}
\end{proposition}

\begin{proof}
(1)$\Leftrightarrow$(2).
Fix $q\neq 0$ and consider the $\F_2$-linear map $\Phi_q\colon V\to\F_4^5$, $H\mapsto Hq$.
Since the trace form on $\F_4^5$ is nondegenerate over $\F_2$, the rank of $\beta_q$ equals $\dim\Phi_q(V)=10-\dim\ker\Phi_q$.
By Lemma~\ref{lem:apn-kernel}, $F$ is APN if and only if $\rank\beta_q=9$ for all $q\neq 0$,
that is, if and only if exactly one nonzero $H\in V$ satisfies $Hq=0$.
This condition depends only on the point $\F_4q$.

(3)$\Leftrightarrow$(4).
We have $F(y)=F(z)$ if and only if $\langle H,yy^*+zz^*\rangle=0$ for all $H\in V$, that is, $yy^*+zz^*\in C$.
Similarly $F(y)=0$ if and only if $yy^*\in C$.
Over $\F_4$, the Hermitian matrices of rank one are the $yy^*$, and those of rank two are the $yy^*+zz^*$ with $y,z$ independent,
since Hermitian forms of a given rank are congruent.
Since $F(\omega y)=F(y)$, every nonzero fiber has at least three elements.
So (3) says that $C$ has no element of rank one or two.

(1)$\Leftrightarrow$(3) is the criterion of Budaghyan, Ivkovic and Kaleyski for triplicate functions~\cite[Thm.~1]{DBLP:journals/ccds/BudaghyanIK23}.
Its hypotheses hold: $F(\omega x)=F(x)$ holds exactly, $F(0)=0$, and $F$ is plateaued, as every quadratic function is.
\end{proof}

Proposition~\ref{prop:herm} links type~64 to Hermitian rank-metric codes.
Here $C$ is an additive code of $2^{15}$ Hermitian matrices with minimum rank distance~3.
It meets the bound of Schmidt for additive Hermitian codes with equality~\cite[Theorem~1]{Schmidt18}.
Since the distance~3 is odd, the rank distribution of $C$ is determined by these parameters~\cite[Theorems~1 and~3]{Schmidt18}.
By duality, so is the rank distribution of $V$; in particular every nonzero element of $V$ has rank at least four.
Then Proposition~\ref{prop:herm}(2) gives the rest: a matrix of rank four annihilates exactly one point of $\PG(4,4)$ and a matrix of rank five none.
So $V$ has 341 elements of rank~4, one for each point, and $1023-341=682$ of rank~5.
The code $C$ has 1,705 elements of rank~3, 12,958 of rank~4 and 18,104 of rank~5.
We also computed these numbers directly, by listing all elements of $V$ and $C$, for the 16 known classes~\cite{artifact}.
Since every nonzero element of $V$ has rank at least four, the graphs $\{(x,Hx):x\in\F_4^5\}$, $H\in V$, pairwise meet in at most a point.
In geometric terms, they are 1,024 generators of the Hermitian polar space $H(9,4)$ with this property~\cite{GowLSV14}.
Here $H(9,4)$ is defined on $\F_4^{10}=\F_4^5\times\F_4^5$ by the Hermitian form $((x,y),(x',y'))\mapsto x^*y'+y^*x'$; the graph of a Hermitian $H$ is totally isotropic for it.
Trombetti and Zullo study maximum additive Hermitian codes~\cite{TrombettiZ21}.
We found no classification of these codes for $5\times 5$ matrices over $\F_4$.
To our knowledge, the realization of type~64 as subspaces of $\Herm_5(\F_4)$ and the characterization of APN by the dual code $C$ (Proposition~\ref{prop:herm}) are new.
The triplicate functions themselves and their 3-to-1 criterion are due to~\cite{DBLP:journals/ccds/BudaghyanIK23}.

So type~64 is the same as the following problem.
\emph{Classify the 10-dimensional $\F_2$-subspaces $V$ of $\Herm_5(\F_4)$ such that every point of $\PG(4,4)$ is annihilated by exactly one nonzero element of $V$.}
The group $\GammaL(5,4)$ acts on these subspaces by $H\mapsto P^*H^\sigma P$.
Equivalent subspaces give EA-equivalent functions.
The converse is not clear, since an EA-equivalence need not respect the $\F_4$-structure.

\subsection{What is known in type 64}
Sixteen of the 29 classes of Theorem~\ref{thm:main} lie in type~64.
They are the Gold functions $x^3$ and $x^9$, two F3 classes, two F4 classes, five F13 classes,
and the triplicate functions \#1, \#3, \#4, \#5 and \#6 of~\cite{DBLP:journals/ccds/BudaghyanIK23}.
The triplicate function \#2 of~\cite{DBLP:journals/ccds/BudaghyanIK23} is EA-equivalent to an F13 class;
we checked an explicit certificate.
All 16 classes also lie in type~73.

We explored type~64 with three search strategies.
They are evidence, not proofs, and they are weaker than the searches of Sections~\ref{sec:search} and~\ref{sec:special} in one respect.
For speed, each search keeps only the first APN function it finds with a given \emph{fingerprint}, the differential spectrum of the ortho-derivative stored as a 64-bit hash, and discards the others.
The kept functions are then named by comparing their ortho-derivative invariants~\cite{DBLP:journals/tit/CanteautCP22} with those of the 29 known classes; no equivalence certificate is computed.
The fingerprint is not a complete invariant.
So what the searches show is that every APN function they found has the fingerprint of one of the 16 known classes of type~64.
A new class with the same fingerprint as a known one would not be detected.
\begin{itemize}
\item \emph{Sparse polynomials, exhaustively.}
We listed all polynomials $x^b+\sum_{k=1}^{t}c_kx^{e_k}$, where the $e_k$ are $t$ distinct exponents among the 24 exponents $2^i+2^j\neq b$ with $j-i$ odd, and the $c_k$ are nonzero.
We ran $b=3$ with $c_k\in\F_4$ and $5\le t\le 10$, $b=3$ with $c_k\in\F_{32}$ and $1\le t\le 4$, and $b\in\{9,33\}$ with $c_k\in\F_4$ and $5\le t\le 9$.
The number of candidates is $\sum\binom{24}{t}(q-1)^t=219{,}963{,}317{,}255$ over these cases.
This extends the search of~\cite[Sec.~9.2]{DBLP:journals/ccds/BudaghyanIK23} to more terms and to $\F_{32}$.
It found $37{,}044{,}315$ APN functions, with the fingerprints of exactly the 16 classes.
\item \emph{Switching in the Hermitian picture.}
For each of the 16 classes we took one space $V$.
For each of its 174,251 subspaces $V_8$ of dimension~8, we computed by linear algebra all spaces $V'\neq V$ of dimension~10 that contain $V_8$ and satisfy Proposition~\ref{prop:herm}(2).
This step has no cap.
The program excludes $V$ itself, which is a completion of every $V_8$.
It computes the fingerprint of every completion it finds.
An independent program recounts the completions on sampled subspaces $V_8$, and checks each of them to be APN with the full difference table~\cite{artifact}.
This includes every switching of codimension one.
The search found 529,903 pairs $(V_8,V')$.
All of them have the fingerprints of 7 of the 16 classes.
Since EA-equivalence need not respect the $\F_4$-structure, one class may contain several inequivalent spaces $V$; we switched only from one space per class.
\item \emph{Random sparse polynomials.}
Type~64 is spanned by the 25 monomials $x^{2^j(2^k+1)}$ with $k$ odd.
More than $4\cdot10^8$ random polynomials with 2 to 14 of these terms, and coefficients in $\F_4$, $\F_{32}$ or $\F_{1024}$,
give the fingerprints of the 16 classes again, and no other fingerprint.
\end{itemize}

\subsection{Why type 64 is hard}\label{sec:t64hard}
\begin{itemize}
\item The generic search of Section~\ref{sec:search} has a Knuth estimate of about $4.5\cdot 10^{57}$ nodes on type~64.
\item A dedicated solver builds $V$ point by point in $\PG(4,4)$, with one matrix of rank four fixed.
Its tree has about $4\cdot 10^{28}$ nodes, and about $3\cdot 10^{27}$ when it branches on the most constrained point.
The constraints prune only at the last two of ten levels.
\item The estimates suggest that isomorph rejection alone is not enough.
The group $\GammaL(5,4)$ has order about $1.5\cdot 10^{15}$.
The Knuth estimate of the tree at level eight is about $4\cdot 10^{28}$ partial spaces.
Dividing by the group order suggests that, even with perfect isomorph rejection, at least about $3\cdot 10^{13}$ inequivalent partial spaces remain there.
This is an estimate, not a proven bound: the nodes of the tree need not be distinct partial spaces, and the set of explored nodes need not be invariant under the group.
\end{itemize}
In short, valid spaces $V$ are rare among partial ones, and the constraints act only at the end.

The evidence for the following conjecture is the three searches above.
None of them is exhaustive on type~64.

\begin{conjecture}\label{conj:t64}
Every quadratic APN function of $\F_2^{10}$ with a self-equivalence of type~64 is EA-equivalent to one of the 16 known classes.
Equivalently, every quadratic canonical triplicate APN function of $\F_2^{10}$ is EA-equivalent to one of these 16 classes.
\end{conjecture}

Since all 16 classes occur in type~73, Conjecture~\ref{conj:t64} and Theorem~\ref{thm:main} together would give the following:
\emph{a quadratic APN function of $\F_2^{10}$ has a nontrivial linear self-equivalence if and only if it is EA-equivalent to one of the 29 classes.}

\section{Results}\label{sec:results}

\subsection{How the 299 types are settled}
Table~\ref{tab:settled} shows how each type is settled.
Most types are settled twice, by a proof and by a search.
The per-type table, with the certificate of each proof and the cost of each search, is in the artifact~\cite{artifact}.

\begin{table}[t]
\centering
\caption{How the 299 types are settled.}\label{tab:settled}
\begin{tabular}{@{}lrl@{}}
\toprule
Settled by & types & remarks \\
\midrule
proof and search & 242 & no APN function \\
proof only & 38 & search too large or not run; no APN function \\
search only & 18 & 9 empty, 9 with APN functions \\
open & 1 & type 64 \\
\midrule
total & 299 & \\
\bottomrule
\end{tabular}
\end{table}

The types settled by search only are the following.
Types~47, 116, 132, 198, 207, 208, 209, 235 and~255 contain no APN function.
Types~24, 73, 129, 170, 171, 175, 176, 180 and~231 contain APN functions.
No other type contains an APN function, except possibly type~64.

\subsection{The 29 classes}

\begin{table}[t]
\centering
\caption{The 29 quadratic APN classes of $\F_2^{10}$ and the types that contain them.
Names of the published classes follow the artifact~\cite{artifact};
BIK refers to~\cite{DBLP:journals/ccds/BudaghyanIK23}.
N1--N4 are the new classes of Table~\ref{tab:new}.}\label{tab:classes}
\begin{tabular}{@{}lcl@{}}
\toprule
Class & $M$ & types \\
\midrule
Gold $x^3$ & 5 & 24, 64, 73, 129, 175 \\
Gold $x^9$ & 5 & 24, 64, 73, 129, 180 \\
F4 $a=1$ ($x^3+\Tr(x^9)$) & 5 & 24, 64, 73 \\
F4 $a=\omega$ & 5 & 64, 73 \\
F3 $i=1$ & 4 & 64, 73, 170 \\
F3 $i=2$ & 4 & 64, 73, 171 \\
F12, two classes & 3, 4 & 73, 176 \\
F12, four classes & 3, 4, 3, 4 & 176 \\
F13 $k=0$, four classes & 5, 5, 4, 3 & 64, 73, 176 \\
F13 $k=2$ & 3 & 64, 73, 129 \\
Beierle--Leander \#1 & 3 & 73, 231 \\
Beierle--Leander \#3 & 5 & 176 \\
Beierle--Leander \#5 & 4 & 73, 129 \\
BIK triplicate \#1, \#3, \#4, \#5, \#6 & 4 & 64, 73 \\
N1 & 3 & 73 \\
N2, N3, N4 & 4 & 73 \\
\bottomrule
\end{tabular}
\end{table}

Table~\ref{tab:classes} lists the 29 classes and the types that contain them.
Every class lies in at least one type other than type~64.
The published classes are the 25 classes of~\cite{DBLP:journals/ffa/BudaghyanCV20,DBLP:journals/tit/BeierleL22,DBLP:journals/ccds/BudaghyanIK23,HellesethIK26},
as collected in~\cite{Bellini26ccz}, with an explicit representative and an equivalence certificate for each.
When we say ``published classes'', we mean these 25.
For some infinite families, \cite{Bellini26ccz} checked only sampled members in dimension ten, so we do not claim that every member of every published family is among them.
The recent exhaustive searches of Ichanska and Kaleyski for extended biprojective functions in dimension ten find only Carlet's classes and the classes of~\cite{HellesethIK26}~\cite[Sec.~7.1]{IchanskaK26}.
These are among the 25.

\begin{proof}[Proof of Theorem~\ref{thm:main}]
Let $F$ have a self-equivalence of prime order of type $t\neq 64$.
If $t$ is one of the 280 types of Section~\ref{sec:lemmas}, then $F$ is not APN.
The same holds for the nine empty types settled by search, including type~47 (Theorem~\ref{thm:t47}).
Otherwise $t$ is one of the nine types listed above.
The searches of Section~\ref{sec:search} and Theorems~\ref{thm:t73} and~\ref{thm:t24} list all APN functions of these types.
Each of them is EA-equivalent to a class of Table~\ref{tab:classes}, by an explicit certificate.
Table~\ref{tab:classes} shows that each class occurs in such a type.
\end{proof}

\subsection{Consequences}
\paragraph*{Every known class has a self-equivalence}
Each of the 25 published classes has a linear self-equivalence of prime order.
Table~\ref{tab:classes} gives, for each class, the types in which it occurs.

\paragraph*{Permutations}
None of the 29 classes is CCZ-equivalent to a permutation.
For the 25 published classes this is shown in~\cite{Bellini26ccz}.
The four new classes have $M<5$ (Table~\ref{tab:new}).
So Theorem~\ref{thm:main} gives the following.

\begin{corollary}
Let $F$ be a quadratic APN function of $\F_2^{10}$ with a linear self-equivalence of prime order, of a type other than~64.
Then $F$ is not CCZ-equivalent to a permutation.
\end{corollary}

\section{Conclusion}\label{sec:conclusion}

We classified the quadratic APN functions of $\F_2^{10}$ with a linear self-equivalence of prime order,
for 298 of the 299 types.
There are 29 classes, four of them new, and none is CCZ-equivalent to a permutation.
Elementary lemmas on the polar form settle 280 types without any search.
The remaining types need exhaustive searches.
The classification and confirmation searches took about 36,400 core-hours.

The open type~64 is a problem on Hermitian matrices.
Classify the 10-dimensional spaces of $5\times 5$ Hermitian matrices over $\F_4$ such that every point of $\PG(4,4)$ is annihilated by exactly one nonzero matrix.
As sets, these spaces are the duals of the maximum additive Hermitian codes of minimum distance three in this size.
We expect that a proof of Conjecture~\ref{conj:t64} needs more structure than our searches use,
for example a geometric argument in the Hermitian polar space $H(9,4)$.

Two questions remain open beyond type~64.
First, the lemmas of Section~\ref{sec:lemmas} hold in every dimension.
However, we do not know whether the searches stay feasible in dimension twelve.
Second, we do not know whether $\F_2^{10}$ has quadratic APN functions without any nontrivial linear self-equivalence.

\appendices
\section{The four new classes}\label{app:new}
Let $g$ be a root of $x^{10}+x^3+1$, so that $\F_{2^{10}}=\F_2(g)$, and let $\omega=g^7+g^6+g^5+g^3+g^2$.
Then $\omega^2+\omega+1=0$, so $\F_4=\{0,1,\omega,\omega^2\}$.
In the artifact, field elements are integers in the basis $1,g,\dots,g^9$, so $\omega$ is 236 and $\omega^2$ is 237.
The following polynomials are representatives of the classes N1--N4 of Table~\ref{tab:new}.
They are the shortest forms we found among the leaves of type~73.
We checked independently, from the full difference table, that each one is APN.
The artifact gives them also as lookup tables, in the file \texttt{data/quadratic\_classes.tsv}~\cite{artifact}.
\begin{align*}
\mathrm{N1}(x)={}&\omega x^{3} + \omega x^{5} + \omega^2 x^{6} + \omega x^{10} + \omega^2 x^{20} + x^{24} + \omega x^{34}\\
&{}+x^{36} + x^{65} + \omega^2 x^{66} + \omega^2 x^{68} + x^{72} + x^{80} + \omega^2 x^{96}\\
&{}+\omega x^{129} + x^{132} + x^{257} + \omega^2 x^{260} + \omega^2 x^{264} + \omega^2 x^{288} + x^{513}\\
&{}+\omega^2 x^{516} + \omega x^{544} + \omega^2 x^{576} + \omega^2 x^{640}
\end{align*}
\begin{align*}
\mathrm{N2}(x)={}&x^{3} + \omega^2 x^{5} + \omega x^{12} + \omega^2 x^{17} + x^{24} + \omega x^{33} + \omega^2 x^{48}\\
&{}+x^{65} + x^{72} + x^{80} + x^{96} + \omega^2 x^{130} + \omega x^{132} + \omega^2 x^{144}\\
&{}+x^{192} + \omega x^{258} + x^{320} + \omega^2 x^{384} + \omega^2 x^{514} + \omega^2 x^{516} + x^{520}\\
&{}+\omega^2 x^{528} + \omega^2 x^{576} + x^{640}
\end{align*}
\begin{align*}
\mathrm{N3}(x)={}&x^{3} + \omega x^{5} + \omega x^{6} + x^{9} + \omega x^{12} + \omega x^{17} + \omega^2 x^{18}\\
&{}+\omega^2 x^{24} + x^{34} + \omega x^{36} + \omega x^{48} + \omega x^{68} + x^{72} + x^{129}\\
&{}+\omega x^{130} + x^{132} + x^{136} + \omega^2 x^{160} + \omega^2 x^{260} + \omega^2 x^{272} + x^{288}\\
&{}+\omega x^{384} + \omega x^{514} + \omega x^{520} + \omega^2 x^{544} + \omega x^{640}
\end{align*}
\begin{align*}
\mathrm{N4}(x)={}&x^{3} + x^{5} + x^{9} + \omega^2 x^{17} + \omega^2 x^{20} + x^{24} + \omega^2 x^{33}\\
&{}+x^{34} + x^{68} + \omega x^{72} + \omega^2 x^{96} + \omega x^{130} + x^{132} + x^{136}\\
&{}+\omega^2 x^{257} + \omega^2 x^{258} + x^{272} + \omega x^{288} + \omega^2 x^{320} + \omega^2 x^{384} + \omega x^{513}\\
&{}+\omega x^{514} + \omega x^{516} + \omega^2 x^{520} + \omega^2 x^{528} + \omega^2 x^{640}
\end{align*}

\section*{Code and data}
The artifact holds the code, the supplied data and the regeneration instructions.
It is available at \artifact.
This paper refers to commit \texttt{6f8a932a9f93}; it will be archived with a DOI at publication.
The artifact contains the matrices of the 299 types, a basis of each space $W(A,B)$, the exclusion certificates with an independent checker, the job lists and completion records of all searches with a checking script, and lookup tables of all 29 classes.
It contains every leaf of types~24 and~73.
The generic search also found APN functions in seven other types, with 4,851,099 leaves in all.
For these, the artifact contains one record per leaf, with its invariant, its class and the result of its equivalence test, but not the lookup tables, which take about 8\,GB.
They are available on request, and the searches that print them take about 9 core-hours.
For type~64, the artifact contains the classified fingerprints, not every function found.
The file \texttt{DATA.md} lists all omitted data.
An equivalence certificate was computed and checked for every leaf of the 298 settled types when the leaves were classified.
The certificates themselves are not stored: the logs record the result, and the quick checks recompute a sample.
For the 4,536 leaves of type~73, the full check takes about 20 minutes on one core.
For the other 4,851,099 leaves it means regenerating the leaves and then about 1,300 core-hours of equivalence tests.
A file of SHA-256 checksums identifies every file.
The README separates three levels: consistency checks of the stored logs, independent checks of the stored certificates and representatives, which take minutes, and the full regeneration of the searches and equivalences.
Field elements are integers in the polynomial basis: $\F_{2^{10}}$ is defined by $x^{10}+x^3+1$, and $\F_{32}$, in the type-24 solver, by $x^5+x^2+1$.

\section*{Acknowledgment}
The author used a large language model, Claude (Anthropic), throughout this work.
It was used to write and run the code of the artifact, to draft and edit the text of all sections, and to check proofs and references.
All results were checked by the author.
For the 298 settled types, every leaf used in the classification was matched to a known class by an explicit EA-equivalence certificate, which was checked directly (Sections~\ref{sec:search} and~\ref{sec:special}).
The exploratory searches of type~64 compute no certificates (Section~\ref{sec:type64}).
The exhaustive searches were validated on cases with a known answer (Sections~\ref{sec:search} and~\ref{sec:special}).
The author takes full responsibility for the content.

\bibliographystyle{IEEEtran}
\bibliography{references}

@inproceedings{DBLP:conf/eurocrypt/Nyberg93,
  author       = {Kaisa Nyberg},
  title        = {Differentially Uniform Mappings for Cryptography},
  booktitle    = {Advances in Cryptology - {EUROCRYPT} '93, Workshop on the Theory and
                  Application of Cryptographic Techniques, Lofthus, Norway, May 23-27,
                  1993, Proceedings},
  series       = {Lecture Notes in Computer Science},
  volume       = {765},
  pages        = {55--64},
  publisher    = {Springer},
  address      = {Berlin, Heidelberg},
  year         = {1993},
  doi          = {10.1007/3-540-48285-7\_6}
}

@article{DBLP:journals/dcc/CarletCZ98,
  author       = {Claude Carlet and
                  Pascale Charpin and
                  Victor A. Zinoviev},
  title        = {Codes, Bent Functions and Permutations Suitable For {DES}-like
                  Cryptosystems},
  journal      = {Des. Codes Cryptogr.},
  volume       = {15},
  number       = {2},
  pages        = {125--156},
  year         = {1998},
  url          = {https://doi.org/10.1023/A:1008344232130},
  doi          = {10.1023/A:1008344232130}
}

@book{DBLP:books/cu/Carlet2021,
  author       = {Claude Carlet},
  title        = {Boolean Functions for Cryptography and Coding Theory},
  publisher    = {Cambridge University Press},
  address      = {Cambridge},
  year         = {2021},
  doi          = {10.1017/9781108606806}
}

@incollection{BrowningDMW10,
  author       = {K. A. Browning and
                  J. F. Dillon and
                  M. T. McQuistan and
                  A. J. Wolfe},
  title        = {An {APN} permutation in dimension six},
  booktitle    = {Finite Fields: Theory and Applications},
  series       = {Contemporary Mathematics},
  volume       = {518},
  pages        = {33--42},
  publisher    = {American Mathematical Society},
  address      = {Providence, RI},
  year         = {2010},
  doi          = {10.1090/conm/518/10194}
}

@article{DBLP:journals/jalc/Yoshiara12,
  author       = {Satoshi Yoshiara},
  title        = {Equivalences of quadratic {APN} functions},
  journal      = {J. Algebraic Combin.},
  volume       = {35},
  number       = {3},
  pages        = {461--475},
  year         = {2012},
  url          = {https://doi.org/10.1007/s10801-011-0309-1},
  doi          = {10.1007/s10801-011-0309-1}
}

@article{DBLP:journals/dcc/YuWL14,
  author       = {Yuyin Yu and
                  Mingsheng Wang and
                  Yongqiang Li},
  title        = {A matrix approach for constructing quadratic {APN} functions},
  journal      = {Des. Codes Cryptogr.},
  volume       = {73},
  number       = {2},
  pages        = {587--600},
  year         = {2014},
  url          = {https://doi.org/10.1007/s10623-014-9955-3},
  doi          = {10.1007/s10623-014-9955-3}
}

@article{DBLP:journals/ffa/BudaghyanCV20,
  author       = {Lilya Budaghyan and
                  Marco Calderini and
                  Irene Villa},
  title        = {On equivalence between known families of quadratic {APN} functions},
  journal      = {Finite Fields Their Appl.},
  volume       = {66},
  year         = {2020},
  url          = {https://doi.org/10.1016/j.ffa.2020.101704},
  doi          = {10.1016/j.ffa.2020.101704},
  note         = {{A}rt. no. 101704}
}

@article{DBLP:journals/ccds/GologluP21,
  author       = {Faruk G{\"{o}}lo{\u{g}}lu and
                  Ji{\v{r}}{\'{\i}} Pavl{\r{u}}},
  title        = {On {CCZ}-inequivalence of some families of almost perfect nonlinear
                  functions to permutations},
  journal      = {Cryptogr. Commun.},
  volume       = {13},
  number       = {3},
  pages        = {377--391},
  year         = {2021},
  url          = {https://doi.org/10.1007/s12095-021-00476-0},
  doi          = {10.1007/s12095-021-00476-0}
}

@article{DBLP:journals/tit/BeierleBL21,
  author       = {Christof Beierle and
                  Marcus Brinkmann and
                  Gregor Leander},
  title        = {Linearly Self-Equivalent {APN} Permutations in Small Dimension},
  journal      = {{IEEE} Trans. Inf. Theory},
  volume       = {67},
  number       = {7},
  pages        = {4863--4875},
  year         = {2021},
  url          = {https://doi.org/10.1109/TIT.2021.3071533},
  doi          = {10.1109/TIT.2021.3071533}
}

@article{DBLP:journals/tit/BeierleL22,
  author       = {Christof Beierle and
                  Gregor Leander},
  title        = {New Instances of Quadratic {APN} Functions},
  journal      = {{IEEE} Trans. Inf. Theory},
  volume       = {68},
  number       = {1},
  pages        = {670--678},
  year         = {2022},
  url          = {https://doi.org/10.1109/TIT.2021.3120698},
  doi          = {10.1109/TIT.2021.3120698}
}

@article{DBLP:journals/ccds/BudaghyanIK23,
  author       = {Lilya Budaghyan and
                  Ivana Ivkovic and
                  Nikolay S. Kaleyski},
  title        = {Triplicate functions},
  journal      = {Cryptogr. Commun.},
  volume       = {15},
  number       = {1},
  pages        = {35--83},
  year         = {2023},
  url          = {https://doi.org/10.1007/s12095-022-00587-2},
  doi          = {10.1007/s12095-022-00587-2}
}

@article{DBLP:journals/ccds/IchanskaBKY26,
  author       = {Nadiia Ichanska and
                  Simon Berg and
                  Nikolay S. Kaleyski and
                  Yuyin Yu},
  title        = {Pushing the {QAM} method for finding {APN} functions further},
  journal      = {Cryptogr. Commun.},
  volume       = {18},
  number       = {3},
  pages        = {727--762},
  year         = {2026},
  url          = {https://doi.org/10.1007/s12095-025-00851-1},
  doi          = {10.1007/s12095-025-00851-1},
  note         = {Preprint: Cryptology ePrint Archive, Paper 2024/1838}
}

@article{YuLIK25,
  author       = {Yuyin Yu and
                  Jingchen Li and
                  Nadiia Ichanska and
                  Nikolay S. Kaleyski},
  title        = {Classification of quadratic {APN} functions with coefficients in
                  {$\mathbb{F}_2$} in dimension 10},
  journal      = {Appl. Algebra Eng. Commun. Comput.},
  volume       = {37},
  number       = {6},
  pages        = {1391--1409},
  year         = {2026},
  url          = {https://doi.org/10.1007/s00200-025-00711-8},
  doi          = {10.1007/s00200-025-00711-8}
}

@article{DBLP:journals/combtheory/GologluK25,
  author       = {Faruk G{\"{o}}lo{\u{g}}lu and
                  Lukas K{\"{o}}lsch},
  title        = {Equivalences of biprojective almost perfect nonlinear functions},
  journal      = {Comb. Theory},
  volume       = {5},
  number       = {3},
  year         = {2025},
  url          = {https://doi.org/10.5070/C65365555},
  doi          = {10.5070/C65365555},
  note         = {{A}rt. no. 7}
}

@article{DBLP:journals/tit/CanteautCP22,
  author       = {Anne Canteaut and
                  Alain Couvreur and
                  L{\'{e}}o Perrin},
  title        = {Recovering or Testing Extended-Affine Equivalence},
  journal      = {{IEEE} Trans. Inf. Theory},
  volume       = {68},
  number       = {9},
  pages        = {6187--6206},
  year         = {2022},
  url          = {https://doi.org/10.1109/TIT.2022.3166692},
  doi          = {10.1109/TIT.2022.3166692}
}

@article{DBLP:journals/corr/abs-2508-04644,
  author       = {Christof Beierle and
                  Philippe Langevin and
                  Gregor Leander and
                  Alexandr A. Polujan and
                  Shahram Rasoolzadeh},
  title        = {Millions of inequivalent quadratic {APN} functions in eight variables},
  journal      = {CoRR},
  volume       = {abs/2508.04644},
  year         = {2025},
  url          = {https://doi.org/10.48550/arXiv.2508.04644},
  doi          = {10.48550/arXiv.2508.04644},
  note         = {Version 2, August 2026}
}

@article{BaudrinBFLNPS25,
  author       = {Jules Baudrin and Christof Beierle and Patrick Felke and Gregor Leander and
                  Patrick Neumann and L{\'{e}}o Perrin and Lukas Stennes},
  title        = {Commutative cryptanalysis as a generalization of differential cryptanalysis},
  journal      = {Des. Codes Cryptogr.},
  volume       = {93},
  number       = {8},
  pages        = {3243--3281},
  year         = {2025},
  doi          = {10.1007/s10623-025-01625-9}
}

@misc{BaudrinCP26,
  author       = {Jules Baudrin and Anne Canteaut and L{\'{e}}o Perrin},
  title        = {Linear self-equivalence of the known families of {APN} functions: a unified point of view},
  howpublished = {Cryptology ePrint Archive, Paper 2026/1012},
  year         = {2026},
  url          = {https://eprint.iacr.org/2026/1012}
}

@misc{HellesethIK26,
  author       = {Tor Helleseth and Nadiia Ichanska and Nikolay Kaleyski},
  title        = {New infinite families of {APN} functions from the switching construction in even dimension},
  howpublished = {Cryptology ePrint Archive, Paper 2026/2401},
  year         = {2026},
  url          = {https://eprint.iacr.org/2026/2401}
}

@misc{Bellini26ccz,
  author       = {Emanuele Bellini},
  title        = {Testing {CCZ}-equivalence to permutations via projections of {W}alsh zero spaces,
                  with applications to {APN} functions},
  howpublished = {Cryptology ePrint Archive, Paper 2026/2800. Submitted to Cryptogr. Commun.},
  year         = {2026},
  url          = {https://eprint.iacr.org/2026/2800}
}

@article{Schmidt18,
  author       = {Kai-Uwe Schmidt},
  title        = {Hermitian rank distance codes},
  journal      = {Des. Codes Cryptogr.},
  volume       = {86},
  number       = {7},
  pages        = {1469--1481},
  year         = {2018},
  doi          = {10.1007/s10623-017-0407-8}
}

@article{TrombettiZ21,
  author       = {Rocco Trombetti and Ferdinando Zullo},
  title        = {On maximum additive {H}ermitian rank-metric codes},
  journal      = {J. Algebraic Combin.},
  volume       = {54},
  number       = {1},
  pages        = {151--171},
  year         = {2021},
  doi          = {10.1007/s10801-020-00973-0}
}

@article{GowLSV14,
  author       = {Rod Gow and Michel Lavrauw and John Sheekey and Fr{\'{e}}d{\'{e}}ric Vanhove},
  title        = {Constant rank-distance sets of {H}ermitian matrices and partial spreads in {H}ermitian polar spaces},
  journal      = {Electron. J. Combin.},
  volume       = {21},
  number       = {1},
  year         = {2014},
  doi          = {10.37236/3534},
  note         = {{A}rt. no. P1.26}
}

@article{Knuth75,
  author       = {Donald E. Knuth},
  title        = {Estimating the efficiency of backtrack programs},
  journal      = {Math. Comp.},
  volume       = {29},
  number       = {129},
  pages        = {121--136},
  year         = {1975},
  doi          = {10.2307/2005469}
}

@article{KolschKK23,
  author       = {Lukas K{\"{o}}lsch and Bj{\"{o}}rn Kriepke and Gohar M. Kyureghyan},
  title        = {Image sets of perfectly nonlinear maps},
  journal      = {Des. Codes Cryptogr.},
  volume       = {91},
  number       = {1},
  pages        = {1--27},
  year         = {2023},
  doi          = {10.1007/s10623-022-01094-4}
}

@article{Wong74,
  author       = {Kai-Tak Wong},
  title        = {The eigenvalue problem {$\lambda Tx + Sx$}},
  journal      = {J. Differential Equations},
  volume       = {16},
  number       = {2},
  pages        = {270--280},
  year         = {1974},
  doi          = {10.1016/0022-0396(74)90014-X}
}

@article{DempwolffE16,
  author       = {Ulrich Dempwolff and Yves Edel},
  title        = {Isomorphisms and automorphisms of extensions of bilinear dimensional dual hyperovals and
                  quadratic {APN} functions},
  journal      = {J. Group Theory},
  volume       = {19},
  number       = {2},
  pages        = {249--322},
  year         = {2016},
  doi          = {10.1515/jgth-2015-0036}
}

@book{LidlN97,
  author       = {Rudolf Lidl and Harald Niederreiter},
  title        = {Finite Fields},
  edition      = {2nd},
  series       = {Encyclopedia of Mathematics and its Applications},
  volume       = {20},
  publisher    = {Cambridge University Press},
  year         = {1997}
}

@misc{Gow17,
  author       = {Rod Gow},
  title        = {Rank-related dimension bounds for subspaces of bilinear forms over finite fields},
  howpublished = {arXiv:1703.07266},
  year         = {2017}
}

@misc{artifact,
  author       = {Emanuele Bellini},
  title        = {Code and data for ``Quadratic {APN} functions of $\mathbb{F}_2^{10}$ with a self-equivalence of prime order''},
  howpublished = {\url{https://github.com/peacker/apn10-self-equivalences}},
  year         = {2026}
}

@article{BrinkmannL08,
  author       = {Marcus Brinkmann and Gregor Leander},
  title        = {On the classification of {APN} functions up to dimension five},
  journal      = {Des. Codes Cryptogr.},
  volume       = {49},
  number       = {1--3},
  pages        = {273--288},
  year         = {2008},
  doi          = {10.1007/s10623-008-9194-6}
}

@inproceedings{Matsui97,
  author       = {Mitsuru Matsui},
  title        = {New block encryption algorithm {MISTY}},
  booktitle    = {Fast Software Encryption ({FSE} 1997)},
  series       = {Lecture Notes in Computer Science},
  volume       = {1267},
  pages        = {54--68},
  publisher    = {Springer},
  year         = {1997},
  doi          = {10.1007/BFb0052334}
}

@misc{IchanskaK26,
  author       = {Nadiia Ichanska and Nikolay S. Kaleyski},
  title        = {Extended biprojective {APN} functions},
  howpublished = {Cryptology ePrint Archive, Paper 2026/2451},
  year         = {2026},
  url          = {https://eprint.iacr.org/2026/2451}
}

@article{IvanyosKQS15,
  author       = {G{\'{a}}bor Ivanyos and Marek Karpinski and Youming Qiao and Miklos Santha},
  title        = {Generalized {W}ong sequences and their applications to {E}dmonds' problems},
  journal      = {J. Comput. Syst. Sci.},
  volume       = {81},
  number       = {7},
  pages        = {1373--1386},
  year         = {2015},
  doi          = {10.1016/j.jcss.2015.04.006}
}

\end{document}